\documentclass[preprint,12pt]{emulateapj}
\usepackage{natbib} 
\usepackage{times}
\usepackage{epsfig,graphicx}
\usepackage{txfonts}

\newcommand{\Msun}{$M_{\odot}$}

\newcommand{\kms}{km s$^{-1}$}
\newcommand{\h}{^h}
\newcommand{\m}{^m}
\newcommand{\s}{^s}
\newcommand{\dg}{^\circ}

\newcommand{\am}{'}

\newcommand{\as}{''}

\begin{document}

\title{Resolved measurements of $X_{\text{CO}}$ in NGC 6946 \\  \textnormal{\today}}
\author{Jennifer Donovan Meyer\altaffilmark{1}, Jin Koda\altaffilmark{1}, Rieko Momose\altaffilmark{2,3}, Masayuki Fukuhara\altaffilmark{2}, Thomas Mooney\altaffilmark{1}, Sarah Towers\altaffilmark{1,4}, Fumi Egusa\altaffilmark{5,6}, Robert Kennicutt\altaffilmark{7}, Nario Kuno\altaffilmark{8,9}, Misty Carty\altaffilmark{10}, Tsuyoshi Sawada\altaffilmark{3,11}, Nick Scoville\altaffilmark{6}}
\altaffiltext{1}{Department of Physics \& Astronomy, Stony Brook University, Stony Brook, NY 11794}
\altaffiltext{2}{Department of Astronomy, University of Tokyo, Hongo, Bunkyo-ku, Tokyo 113-0033, Japan}
\altaffiltext{3}{National Astronomical Observatory of Japan, Mitaka, Tokyo 181-8588, Japan}
\altaffiltext{4}{Department of Physics, Western Michigan University, Kalamazoo, MI 49008}
\altaffiltext{5}{Institute of Space and Astronautical Science, Japan Aerospace Exploration Agency, Chuo-ku, Sagamihara, Kanagawa 252-5210, Japan}
\altaffiltext{6}{Department of Astronomy, California Institute of Technology, Pasadena, CA 91125}
\altaffiltext{7}{Institute of Astronomy, University of Cambridge, Cambridge CB3 0HA, United Kingdom}
\altaffiltext{8}{Nobeyama Radio Observatory, Minamimaki, Minamisaku, Nagano, 384-1305, Japan}
\altaffiltext{9}{The Graduate University for Advanced Studies (SOKENDAI), 2-21-1 Osawa, Mitaka, Tokyo 181-0015}
\altaffiltext{10}{Department of Astronomy, University of Maryland, College Park, MD 20742}
\altaffiltext{11}{Joint ALMA Office, Alonso de C\'ordova 3107, Vitacura, Santiago 763-0355, Chile}

\begin{abstract}
We present the largest sample to date of giant molecular clouds (GMCs) in a substantial spiral galaxy other than the Milky Way. We map the distribution of molecular gas with high resolution and image fidelity within the central 5~kpc of the spiral galaxy NGC 6946 in the $^{12}$CO (J=1-0) transition. By combining observations from the Nobeyama Radio Observatory 45-meter single dish telescope and the Combined Array for Research in Millimeter Astronomy (CARMA) interferometer, we are able to obtain high image fidelity and accurate measurements of L$_{CO}$ compared with previous purely interferometric studies. We resolve individual giant molecular clouds (GMCs), measure their luminosities and virial masses, and derive $X_{\text{CO}}$ -- the conversion factor from CO measurements to H$_{2}$ masses -- within individual clouds. On average, we find that $X_{\text{CO}}$ = 1.2 $\times$ 10$^{20}$ cm$^{-2}$ / (K km s$^{-1}$), which is consistent within our uncertainties with previously derived Galactic values as well as the value we derive for Galactic GMCs above our mass sensitivity limit. The properties of our GMCs are largely consistent with the trends observed for molecular clouds detected in the Milky Way disk, with the exception of six clouds detected within $\sim$400~pc of the center of NGC 6946, which exhibit larger velocity dispersions for a given size and luminosity, as has also been observed at the Galactic center. 
\end{abstract}


\section{Introduction}

Studies of extragalactic giant molecular clouds (GMCs) shape our understanding of spiral galaxies on both small and large scales. Within GMCs, on scales smaller than these massive bound gas structures, individual gas cores inside the clouds condense and form new stars. On scales larger than GMCs, global galaxy dynamics govern the motions of these massive gas clouds, where we study the physical environments which create and destroy them. Much of what is known about the evolution of the interstellar medium in spiral galaxies has been learned by analyzing the properties of these large molecular clouds of gas (i.e., \citealt{Wada08, Koda09, Tasker09, Dobbs11, Egusa2011}).

In order to study the gas of GMCs, which is primarily composed of molecular hydrogen, other molecular tracers -- the most common of these being the lower rotational transitions of CO -- are typically observed since the overall gas temperatures are too low to directly excite H$_{2}$ line emission. Accurate determinations of the masses of GMCs are thus dependent upon an accurate relation between CO flux and H$_{2}$ mass, known in the literature as the $X_{\text{CO}}$ factor. Values of this conversion factor have been derived for molecular clouds within our own Galaxy \citep{Solomon87, Scoville87, Dame01} as well as compiled for nearby galaxies \citep{Bolatto08}; to within a factor of two, the typical value seems to hover around 2-3 $\times$ 10$^{20}$ cm$^{-2}$ / (K km s$^{-1}$) (\citealt{Solomon87}, but see \citealt{Heyer09}; \citealt{Scoville87, Wilson90, Dame01, Oka01, Blitz07, Bolatto08}), with higher values observed in the SMC \citep{Israel97, Leroy07, Blitz07, Bolatto08}. 

The conversion factor, derived from Galactic clouds and resolved extragalactic GMCs in various environments, is used to derive molecular gas masses for unresolved GMCs in galaxies out to high redshift. Deriving this value in nearby GMCs is therefore crucial to our understanding of the amount of molecular gas, molecular fraction, and star formation efficiency (star formation rate per unit gas mass) in galaxies as a function of redshift and environment.

For galaxies other than the Milky Way, GMC studies are typically resolution-limited (see the discussion in \citealt{Rosolowsky06}). In order to identify individual molecular clouds, only nearby galaxies can be studied, which has kept the available sample of extragalactic clouds relatively small. For this reason, investigations into individual extragalactic GMC properties, and variations of these with GMC environments, have been mostly limited to clouds inhabiting Local Group galaxies (e.g., \citealt{Wilson90}; \citealt{Arimoto96}; \citealt{Israel97}; \citealt{Boselli02}; \citealt{Blitz07}; \citealt{Bolatto08}; \citealt{Hughes2010}; \citealt{Bigiel2010}; \citealt{Leroy11}). 

Even when they are resolvable, the determination of GMC boundaries is not straightforward. In addition to being limited by instrumental resolution, the appropriate spatial separation of two (or more) adjacent clouds is not always obvious in crowded or highly blended regions. Various studies of Galactic GMCs (e.g., \citealt{Solomon87, Scoville87}) have solved this issue by identifying clouds ``by hand" above a specific temperature contour (which is dependent upon the sensitivity of the instrument). More recently, automated algorithms have been developed (e.g., \citealt{Williams94, Rosolowsky06}) in order to identify individual clouds; these commonly used codes differ in their clump identification philosophy, particularly with regard to the amount of flux included in identified clumps, as discussed at length by \citet{Rosolowsky06}.

Before even identifying individual GMCs, many studies of extragalactic clouds suffer from the more fundamental problem of missing zero spacing information. Observing the gas distribution of a galaxy with only an interferometer neglects extended flux on the largest scales, where it is ``resolved out" by the beam of an interferometric array. This effect can be quite significant \citep{Koda11}. 

\subsection{CARMA \& NRO45 CO Survey of Nearby Galaxies}

To investigate GMC evolution in galactic disks and resolve the physics which controls the star formation rate within GMCs, we will present resolved observations of GMCs across the disks of nearby galaxies in a CO Survey of Nearby Galaxies based on observations using the CARMA interferometer and the Nobeyama Radio Observatory 45-meter (NRO45) single dish telescopes. To date, 13 galactic disks have been observed with both CARMA and NRO45 as part of the survey, and 4 additional galaxies have been observed with the single dish. The design of the survey will be discussed in Koda et al. (in prep). In our survey, we will resolve individual GMCs in a significant sample of nearby spiral galaxies with a variety of morphologies in order to study the evolution of molecular clouds and star formation.

In this paper, we combine observations from CARMA and NRO45 in order to achieve high resolution, as well as extremely high image fidelity, and resolve individual extragalactic GMCs in NGC 6946 to highlight the results made possible by our survey. In Section II we describe the observations, and we present the results in Section III as well as a discussion of the boundaries of individual GMCs. We utilize the well known $\sc{CLUMPFIND}$ algorithm \citep{Williams94} to derive sizes and velocity dispersions for individual GMCs. In Section IV, we discuss the properties of our detected GMC sample and determine the conversion factor $X_{\text{CO}}$ within individual clouds. We summarize our conclusions in Section V. 

\section{Observations}
\subsection{Nobeyama Radio Observatory}\

We observe NGC 6946 in the $^{12}$CO (1-0) transition with the Nobeyama 45-meter single dish telescope\footnote[12]{Nobeyama Radio Observatory is a branch of the National Astronomical Observatory of Japan, National Institutes of Natural Sciences.}, using the Beam Array Receiver System (BEARS) instrument. Observations were performed during the early months of 2008, 2009, and 2010 (throughout our three year observing program at NRO) as part of our CO survey of nearby galaxies. BEARS is a multi-beam receiver with 25 beams, which are aligned in a 5$\times$5 orientation. The FWHM of the 45-m dish is 15$\as$ at 115~GHz (19.7$\as$ after regridding), and we observe with channel increments of 500 kHz and Hanning smooth for a velocity resolution of 2.54 \kms. After dropping the edge channels, we use a bandwidth of 265 MHz (690 \kms). For most of the scans, the system temperature in the double side band (DSB) of BEARS ranges from 300-400~K; scans with T$_{sys}$ much higher than 400~K (due to being observed during daylight hours) are heavily down-weighted and as a result do not contribute much to the final images. The ratio of the upper-to-lower side band (known as the scale factor) was confirmed each year to be within a few percent by observing Galactic CO sources (e.g., AGB stars) with BEARS and the single side band (SSB) S100 receiver and making subsequent corrections. We use the S40 receiver and Galactic masers for pointing. The pointing was checked roughly every two hours during the observations and was accurate to 2-3$\as$. We convert T$_{A*}$ to T$_{mb}$ assuming that the main beam efficiency of the telescope is 0.4 (i.e., T$_{mb}$ = T$_{A^*}$ / 0.4). 

Using on-the-fly (OTF) mapping, NGC 6946 was scanned in the RA and Dec directions, and positions external to the galaxy (OFF positions) were observed between scans. Extrapolating between OFF scans on opposite sides of the galaxy greatly reduced non-linearities in the spectral baselines. The duration of each scan (ON + OFF) was $\sim$1 minute, and the entire galaxy was mapped in $\sim$40 minutes; a total of 36 usable maps were taken for a total of 24 hours of observation time (including ON, OFF, and slew time). The scans were separated by 5$\as$, resulting in oversampling by a factor of 3 compared to the 15" FWHM of the beam, which is necessary in order to achieve Nyquist sampling (5.96$\as$) of $\lambda_{CO}$/D = 11.92$\as$ (where $\lambda_{CO}$ is the observed wavelength, 2.6 mm, and D is the antenna diameter, 45~m). The data reduction and sky subtraction, performed by interpolating between the OFF scans for each OTF (ON) scan, were completed using the NOSTAR package developed at the Nobeyama Radio Observatory. Spatial baseline-subtracted maps were made separately from the scans in the RA and Dec directions in order to minimize systematic errors in the scan directions, and these were subsequently co-added. The rms noise of these single dish observations is 0.13 K (0.57 Jy beam$^{-1}$). 

\subsection{CARMA}

To complement the single dish observations, NGC 6946 was also observed in the $^{12}$CO (1-0) transition in April 2009 with the C and D configurations of CARMA. CARMA is a 15-element interferometer which combines six 10-meter antennae (originally the Owens Valley Radio Observatory, or OVRO) with nine 6-meter antennae (formerly the Berkeley-Illinois-Maryland Association, or BIMA) to achieve superior $\it{uv}$-coverage compared to either of its predecessors. The observations of NGC 6946 were performed using three dual side bands, each with 63 channels, for a total bandwidth of $\sim$100 MHz (after removing six edge channels per sideband) and a channel width of 2.54 \kms. 
After a total of $\sim$21 hours on source (including calibrators), we achieve an rms of 0.73 K. 

\subsection{Combination procedure}

In order to combine the data from the single dish and interferometer, the NRO45 image is converted to visibilities, combined with the CARMA visibilities in the $\it{uv}$-plane, and the new $\it{uv}$ dataset is imaged together. We follow the procedure thoroughly described in \citet{Koda11} for the imaging of M51, and we refer the reader to that paper for the details of the NRO45 deconvolution, combination in $\it{uv}$-space, and imaging process. In that paper, the relative $\it{uv}$-coverage of the single dish and interferometer observations enabled the single dish visibilities to be flagged beyond 4 k$\lambda$, as CARMA visibilities existed down to this value. We keep the NRO45 visibilities out to 10 k$\lambda$ in order to ensure sufficient overlap between the two sets of $\it{uv}$-coverage. The rms of the combined cube is 0.11 Jy beam$^{-1}$ (1.9 K) using the combined synthesized beam, discussed below. We maintain the instrumental velocity resolution of 2.54 \kms~to optimize our ability to resolve, not only to detect, GMCs. The observing parameters for both the single dish and interferometric data sets are summarized in Table~\ref{table:obs}.

\begin{table}
\begin{center}
\caption{$^{12}$CO observations of NGC 6946 \label{table:obs} }
\begin{tabular}{cccc}
\hline
\hline
 & NRO 45 & CARMA & Combined \\
Observing time (hours) & 24 & 21 & 45 \\ 
Bandwidth (MHz) & 265 & 100 & 100 \\
Velocity resolution (\kms) & 2.54 & 2.54 & 2.54 \\
Beam size ($\as$) & 19.7 & ... & 2.3 $\times$ 2.3 \\
RMS (K) & 0.13 & 0.73 & 1.9 (0.11 Jy beam$^{-1}$) \\
\hline
\end{tabular}
\end{center}
\end{table}

\subsubsection{Combined synthesized beam}

Combining interferometric and single dish information is non-trivial, as each is imprinted with the intrinsic beam size with which the observation is made. Typically, when deconvolving interferometric data, the process (e.g., CLEAN) removes the synthesized beam pattern at the location of each emission peak detected in the dirty map above some threshold and replaces it with a convolution beam, which is typically Gaussian in shape. The convolution beam in the case of a combined (interferometer + single dish) data set requires a beam comprised of weighted components from both telescopes and is effectively a superposition of two Gaussians with quite different FWHM sizes. Cleaning algorithms, such as $\it{invert}$ within the interferometric data reduction package $\it{miriad}$, yield an output dirty beam whose point spread function can be fitted with one Gaussian to achieve the beam size, but we find that this method does not lead to flux conservation since the two-component beam does not have a simple Gaussian shape. 

In order to ensure that we conserve the flux in our combined NRO45+CARMA cube, we assume that the single dish observations accurately measure the total flux of NGC 6946. We calculate the flux in the single dish dirty map and compare it to the flux measured in the combined cleaned map; if the flux is conserved, these should be equal. We choose the beam size which leads to flux conservation by setting equal the single dish flux per single dish beam (with units Jy beam$_{SD}^{-1}$), both of which we know, and the flux in the combined dirty map per combined beam (with units Jy beam$_{comb}^{-1}$), where we know the flux but not the beam size:

\begin{equation} 
\frac {flux_{SD} (Jy) } {beam_{SD} (\as)} = \frac {flux_{combined} (Jy) } {beam_{combined}(\as)}.
\end{equation}

In this way, we calculate the intrinsic size of the combined beam (2.3$\as$ $\times$ 2.3$\as$) and use this value to clean the combined map. 
We refer the reader to \citet{Koda11} for a complete discussion of this step in the data combination procedure.

\section{Results}

\subsection{Imaging}

\begin{figure*}
\plottwo{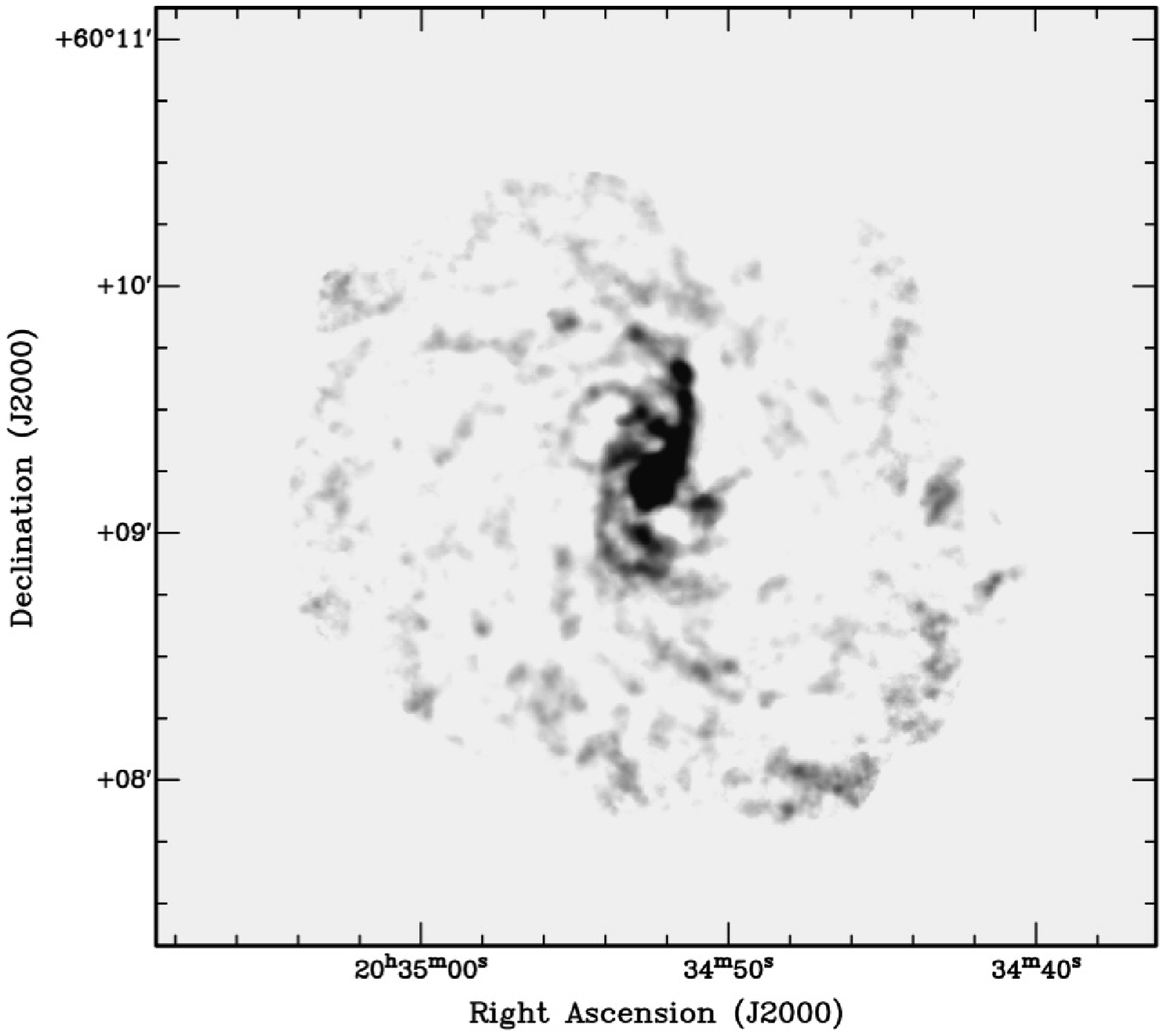}{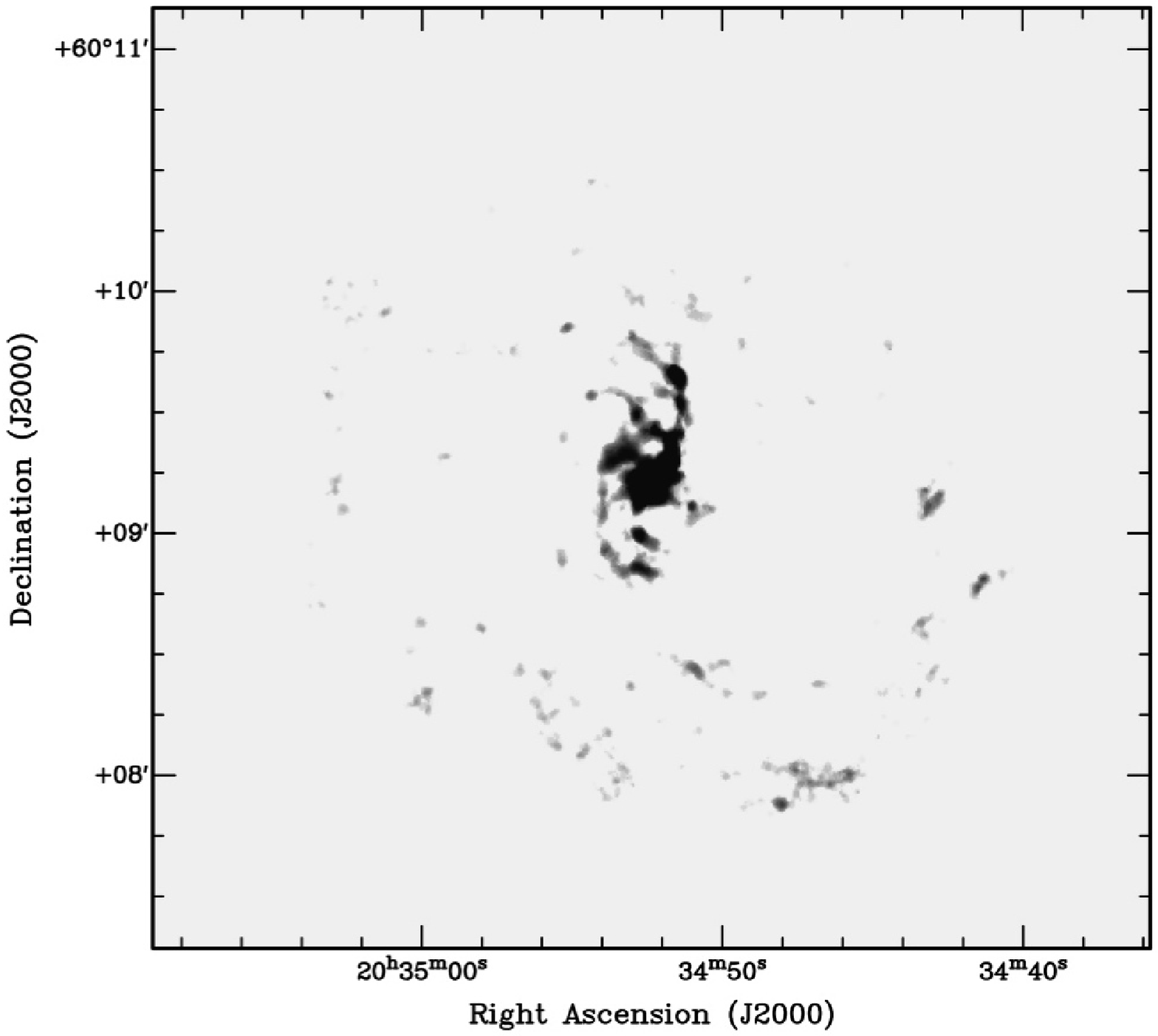}
\caption{CO map of NGC 6946 (left) and flux assigned to 134 clumps by $\sc{CLUMPFIND}$ (right). Our imaging is optimized to resolve GMCs but not to detect less massive GMCs. 
\label{co} }
\end{figure*}

\begin{figure*}
\plotone{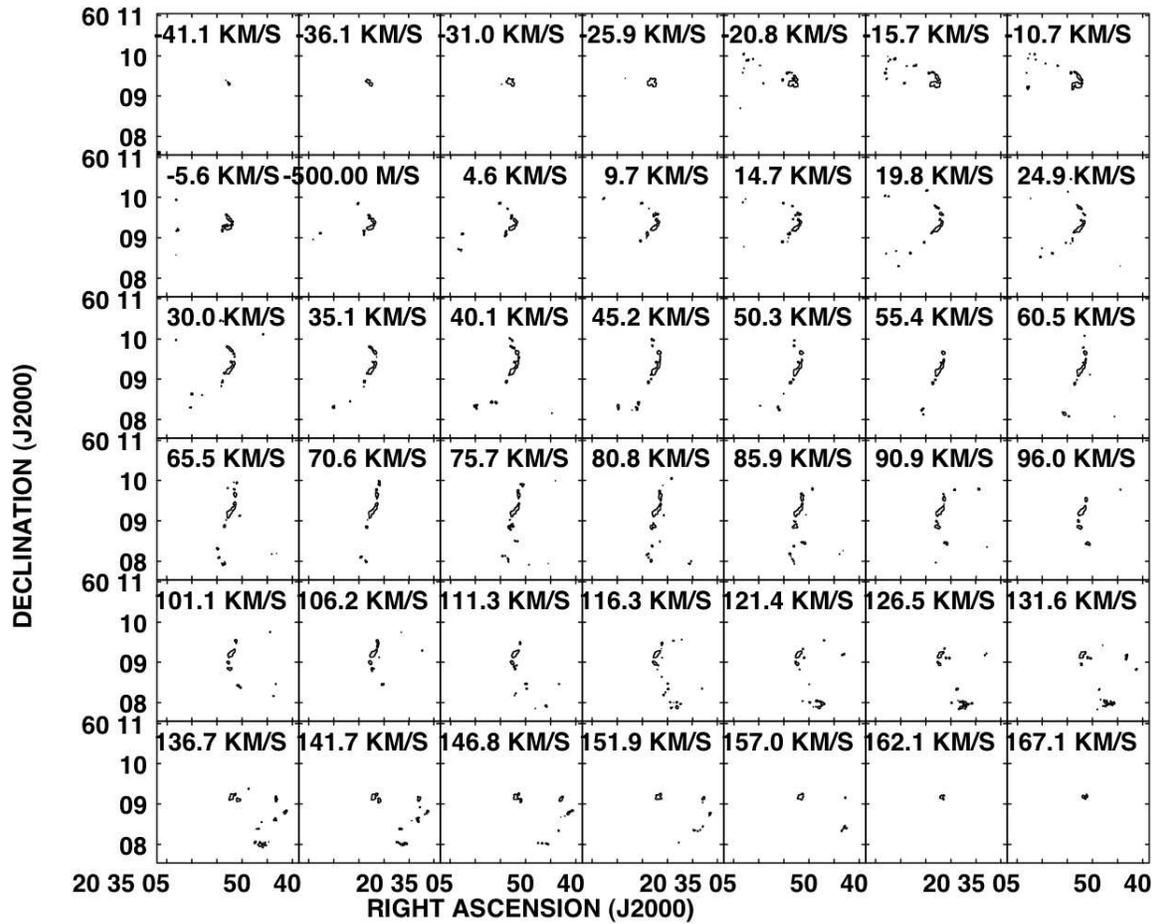}
\caption{CO emission from NGC 6946; every other velocity channel is shown. The contour plotted represents 2$\sigma$ (0.22 Jy beam$^{-1}$).  \label{chanmap} }
\end{figure*}

NGC 6946 is a well-known nearby spiral galaxy. Its general properties are listed in Table~\ref{table:properties}, and our CO imaging is presented in Figures~\ref{co} and \ref{chanmap}. The combined $\it{uv}$ data sets are imaged with purely uniform weighting (robust = -5), where a weighting scheme is applied to the visibilities which is inversely proportional to the sampling density (in effect, weighting all cells the same at the cost of higher $\it{rms}$ noise). This method takes advantage of the highest possible resolution of the array and yields a beam size of 2.3$\as$ (calculated as described in Section 2.3.1). 

\begin{figure*}
\plottwo{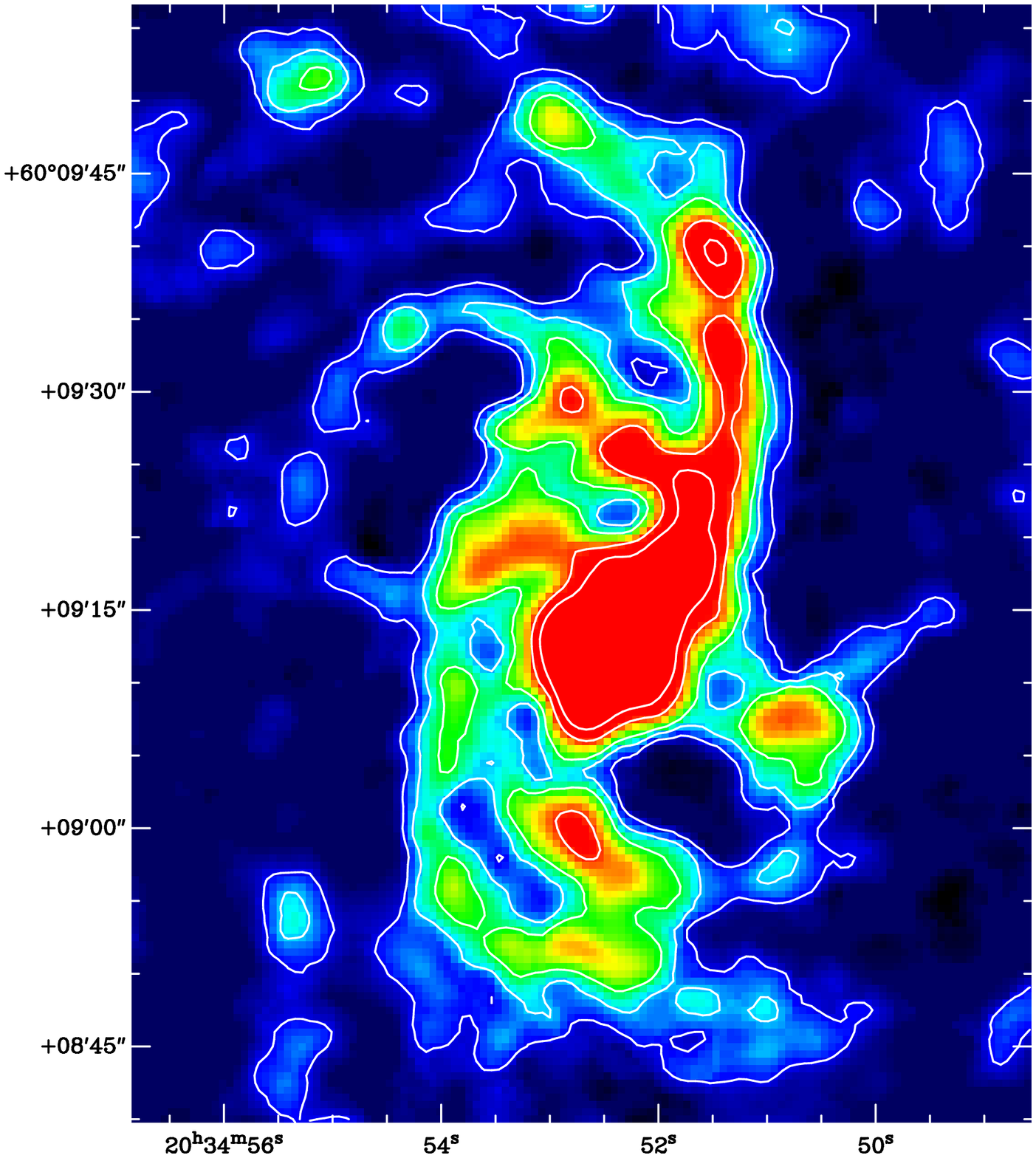}{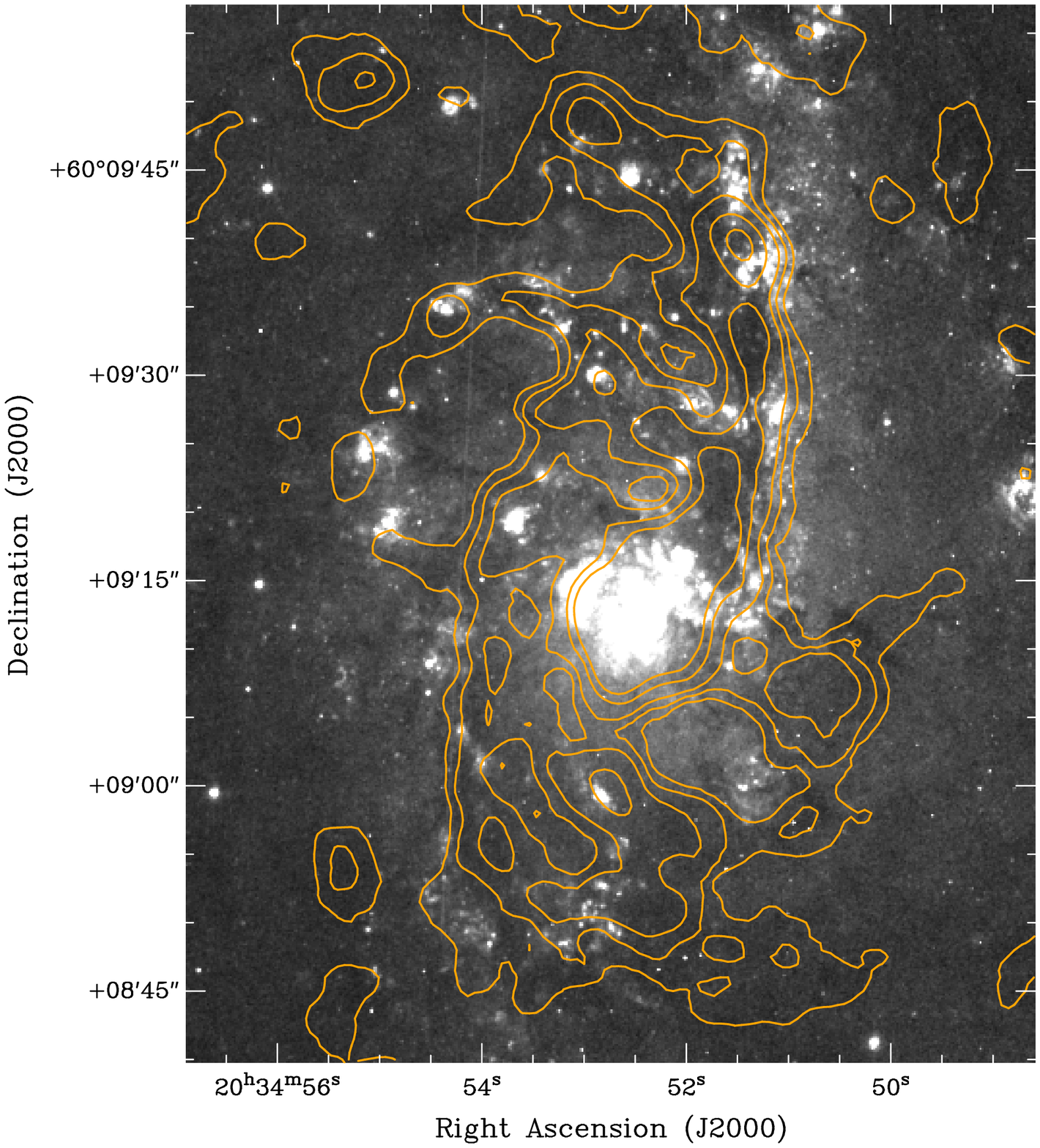}
\caption{The central region of NGC 6946 is shown. CO contours are overlaid on the CO imaging (left) and H$\alpha$ imaging from HST (right). The molecular gas coincides with the spiral arm structures, namely the apparent stellar features and dust lanes. North is up, and east is to the left. \label{halpha} }
\end{figure*}

We overlay the high resolution CO imaging on a Hubble Space Telescope (HST) H$\alpha$ image of the center of NGC 6946 in Figure~\ref{halpha}. The CO emission traces the spiral arm and bar structures at the center of the galaxy. To the north and southeast of the nucleus, the CO emission and dust lanes apparent in the H$\alpha$ image appear to be well matched. As the spiral arm structure consists of adjacent stellar light and dust features (inferred from the H$\alpha$ absorption), where the CO does not match the absorption precisely, its peaks tend to be found on the edge of the absorption feature nearer to the stellar component. 

\begin{figure*}
\plottwo{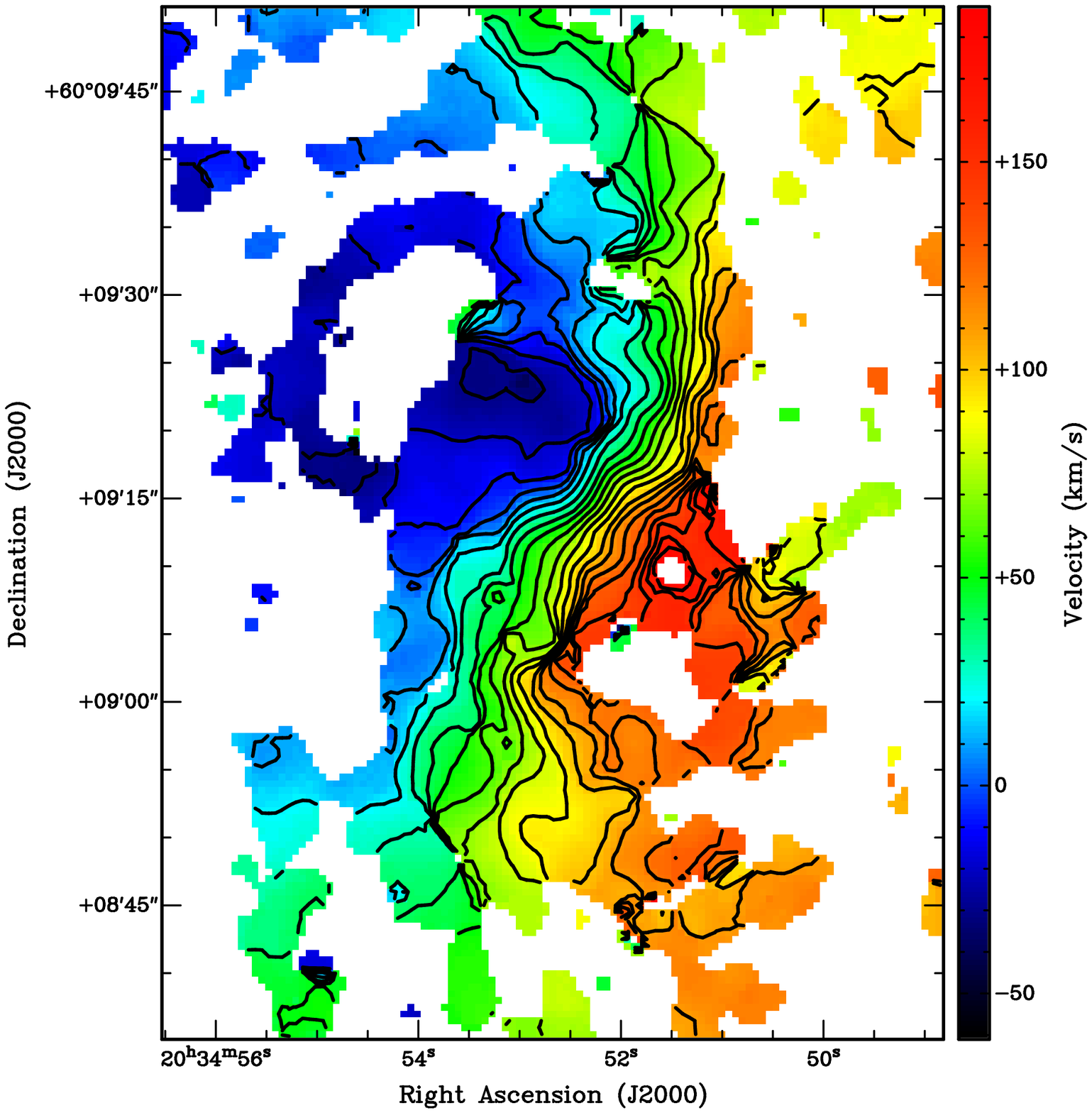}{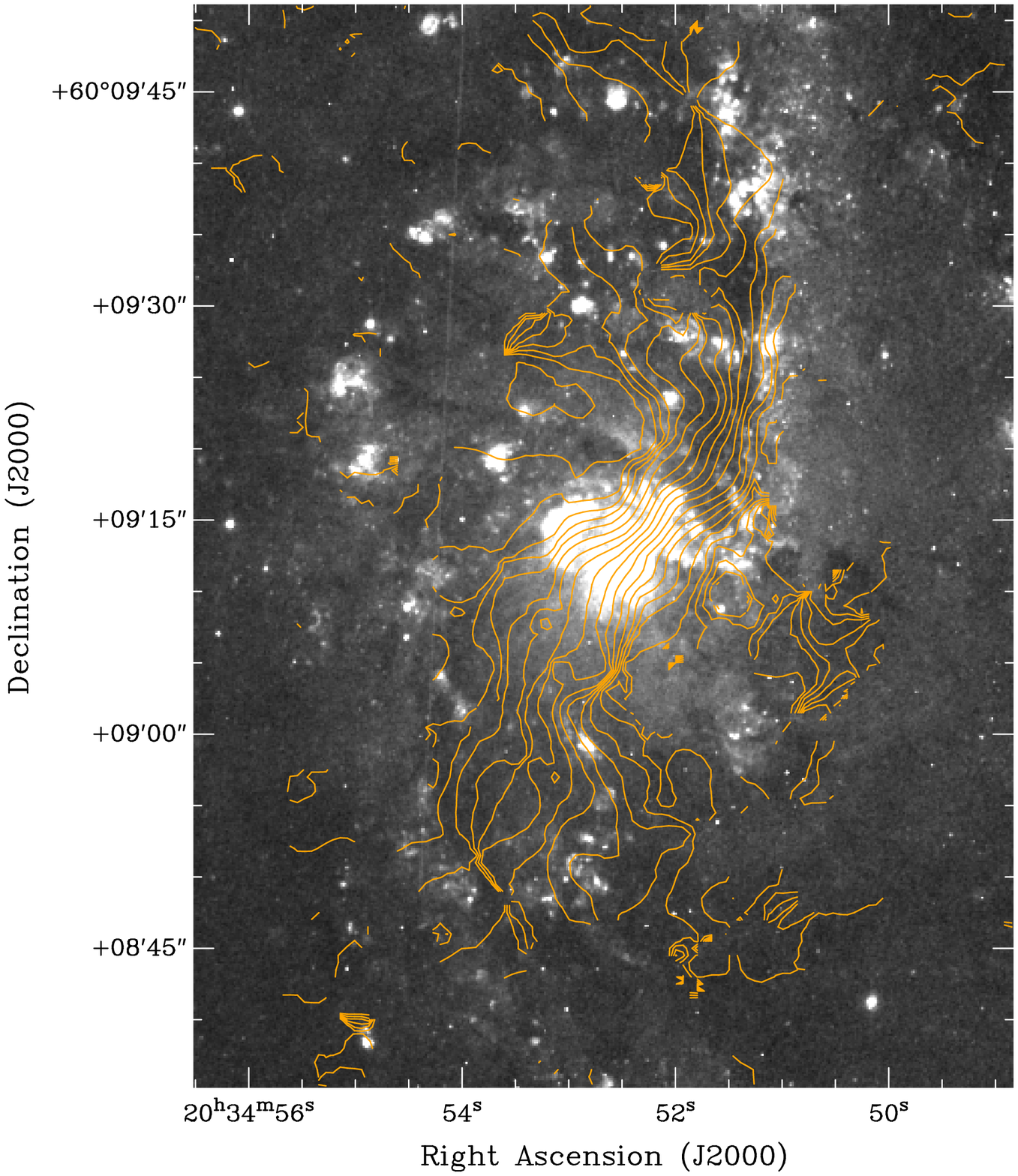}
\caption{Velocity field of the CO emission in NGC 6946. CO velocities are color-coded (left) with contours drawn every 10 \kms. The same contours are shown overlaid on the HST H$\alpha$ image (right). \label{velfield} }
\end{figure*}

NGC 6946 is known to host a molecular bar \citep{Ball85}, and we show the velocity field of NGC 6946 in Figure~\ref{velfield} to accentuate the regions where narrow spatial regions span large ranges in velocity. These are the offset ridges of the galactic bar, which largely coincide with the edges of the dust lanes seen in the H$\alpha$ image.


\begin{table}[t]
\begin{center}
\caption{Properties of NGC 6946  \label{table:properties} }
\smallskip
\begin{tabular}{cc}
\hline
\hline
RA$_{J2000}$ & 20$\h$34$\m$52.3$\s$ \\
Dec$_{J2000}$ & 60$\dg$09$\am$14$\as$ \\
Morphology$^{a}$ & SABcd \\
Distance$^{b}$ & 6.08 Mpc \\
Optical velocity$^{a}$ & 40 \kms \\
Major diameter$^{a}$ & 11.5' \\
\hline
\end{tabular}
\end{center}
\centerline{REFERENCES: $\it{a}$: NASA/IPAC Extragalactic Database (NED);} 
\centerline{$\it{b}$: \citealt{Herrmann08}.}
\end{table}

\subsection{Defining giant molecular clouds}

Given the distance of NGC 6946 and our spatial resolution of 68~pc, we expect to be sensitive to the largest GMCs and giant molecular associations (GMAs). The typical size of a Galactic GMC is 40~pc \citep{ScovilleSanders87}, but measurements range from less than 20~pc to over 100~pc \citep{Solomon87, Scoville87}. We utilize $\sc{CLUMPFIND}$ \citep{Williams94}, a well-known algorithm designed to search image cubes in three dimensional (x-y-v) space for coherent emission, in order to decompose our detected CO emission into individual molecular clouds (e.g., \citealt{Koda09, Egusa2011}). 

\begin{figure}
\includegraphics[width=1in]{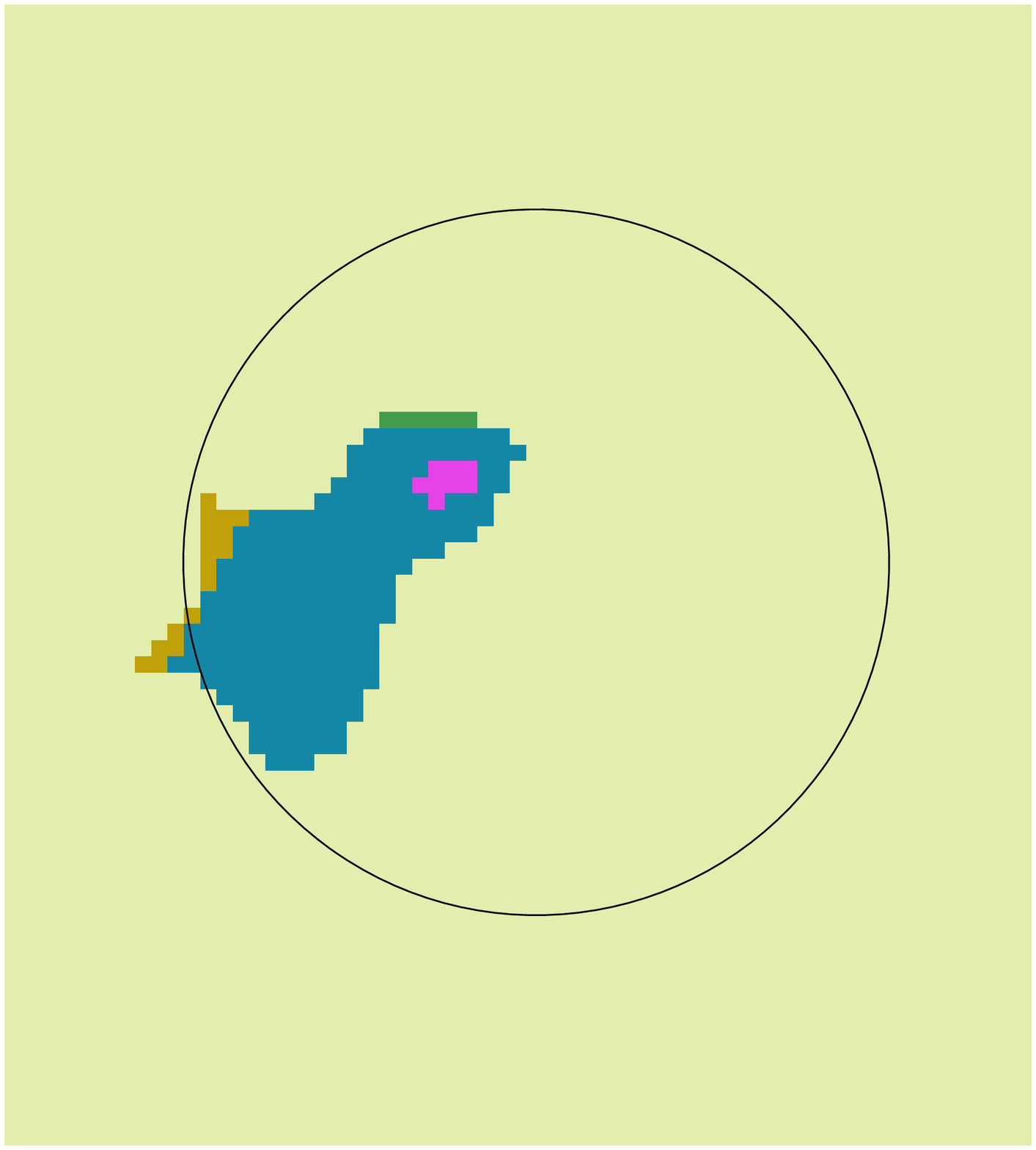}
\includegraphics[width=1in]{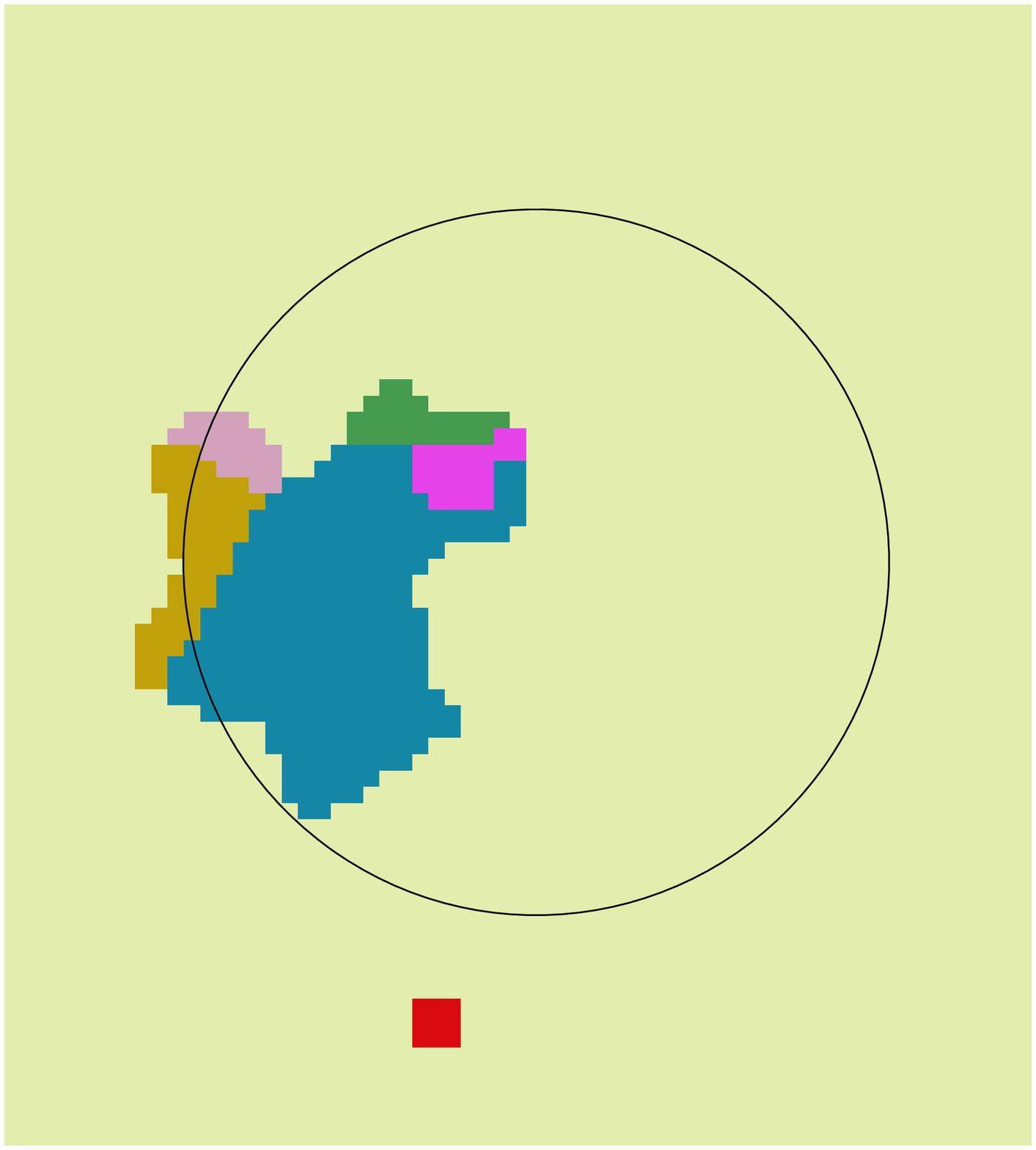}
\includegraphics[width=1in]{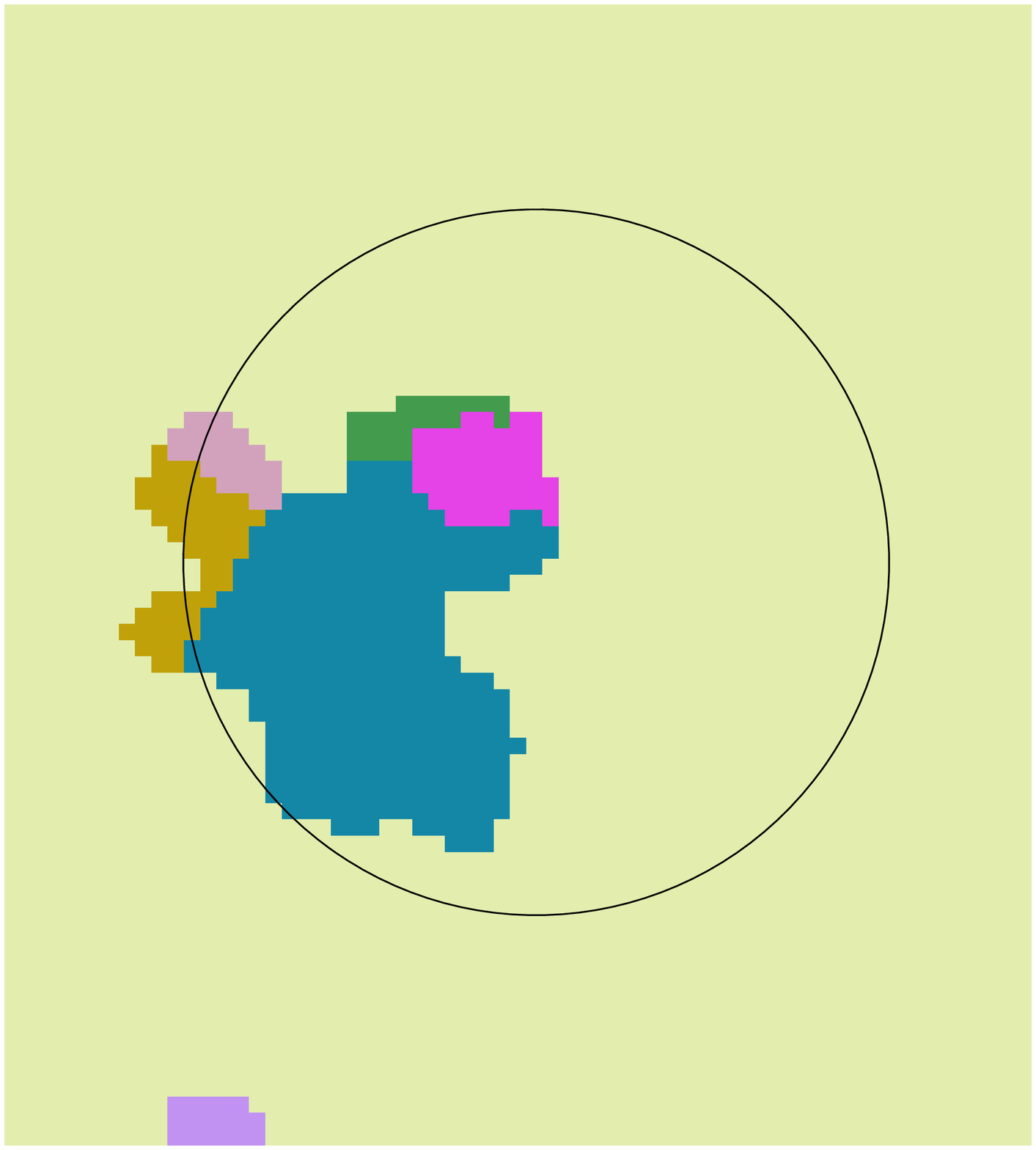} \\
\includegraphics[width=1in]{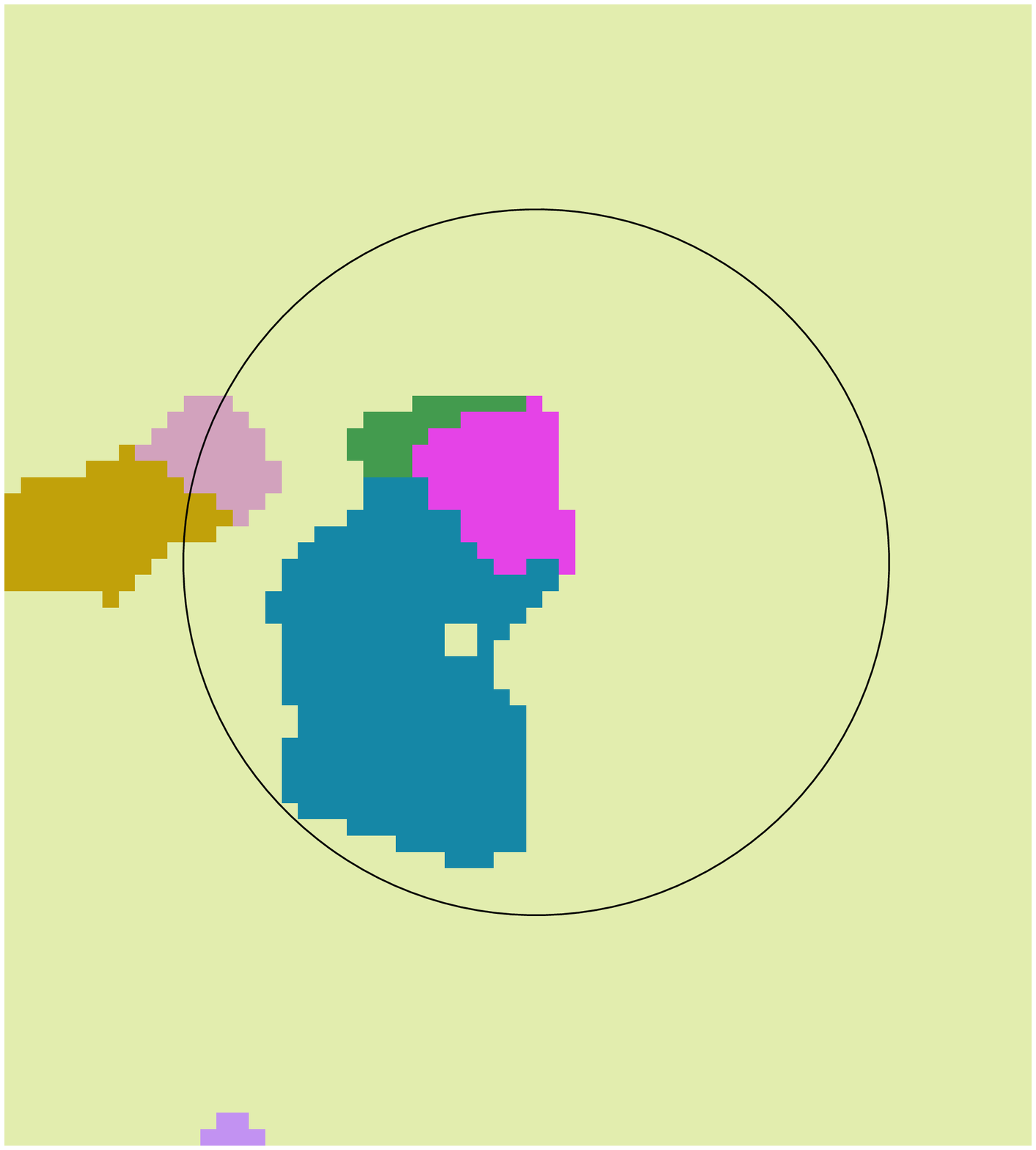} 
\includegraphics[width=1in]{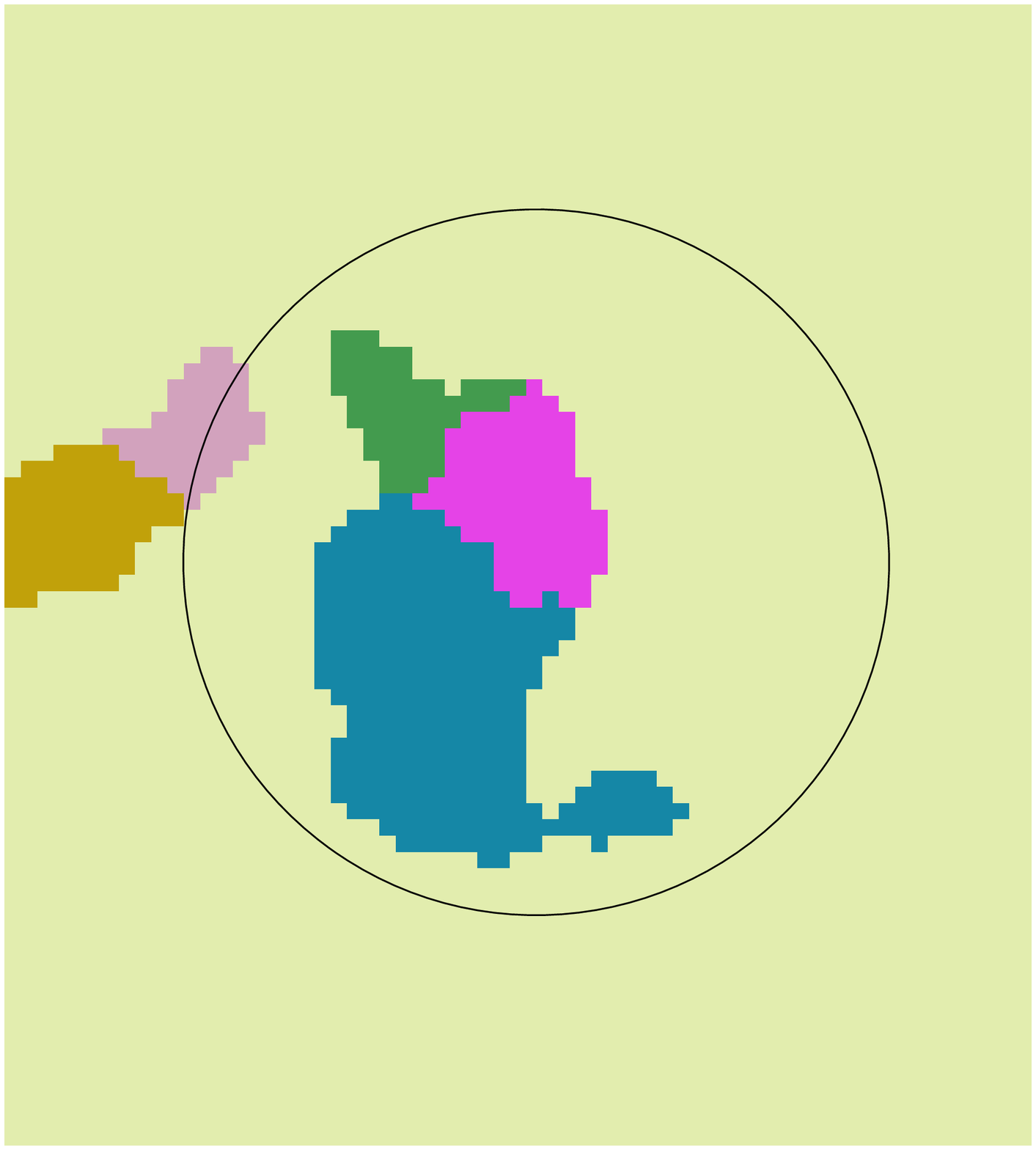} 
\includegraphics[width=1in]{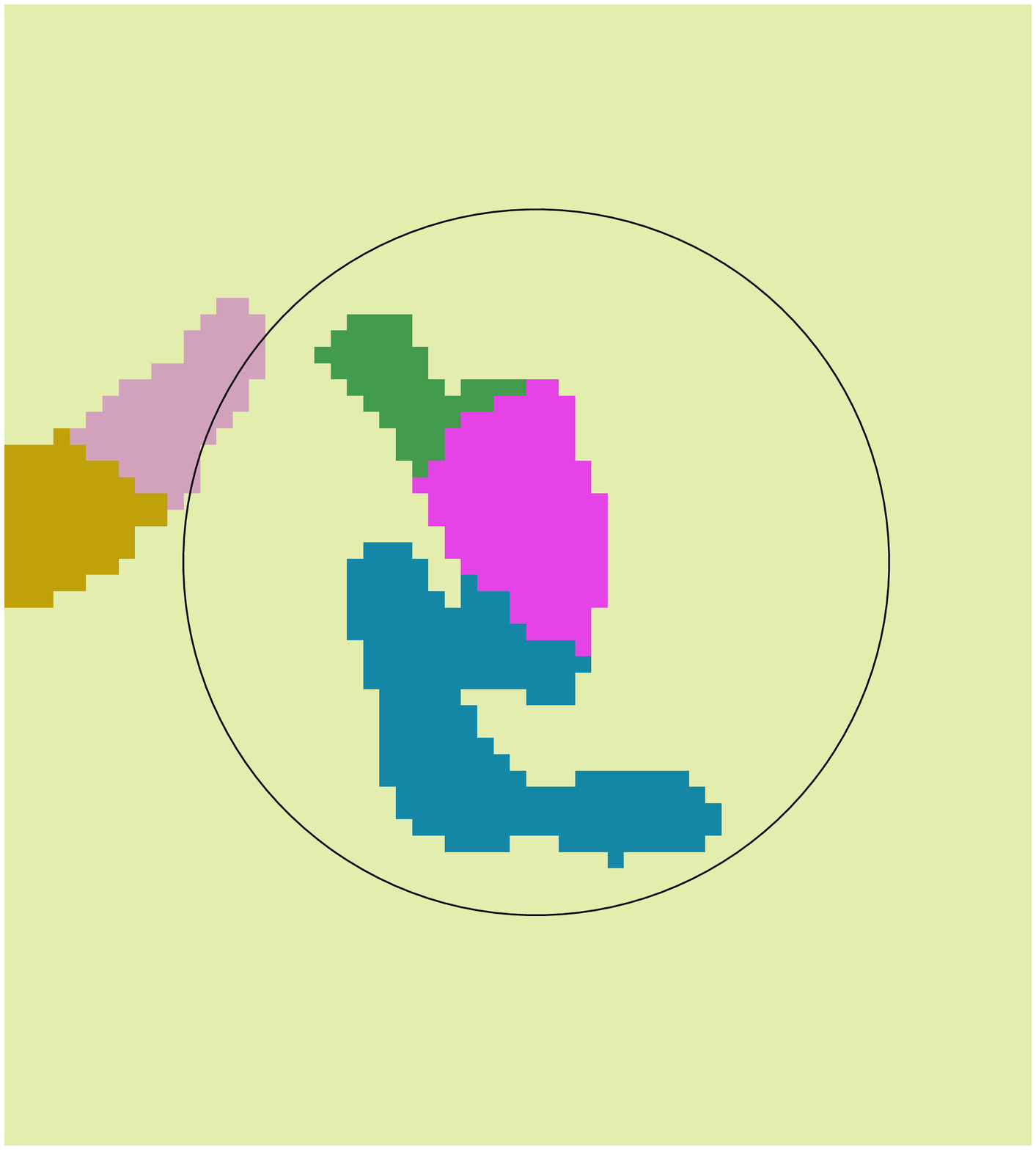}
\caption{Clouds detected in the central region of NGC 6946. Channels are separated by 5 \kms, and colors are coded by clump number. The black circles indicate the same scale on each panel. \label{cloud1} }
\end{figure}

$\sc{CLUMPFIND}$ assigns emission to individual clouds by contouring the emission in a data cube at multiples of the rms noise, identifying peaks, and following the peaks to lower intensities. We find $\sc{CLUMPFIND}$ to be the simplest algorithm for cloud decomposition, as it requires only two user-specified inputs: the contour increment and the minimum contour level (i.e., where to stop), and it assumes no physical priors in the clump decomposition process. We find that the most believable output is returned when using twice the rms noise of the data cube (0.11 Jy beam$^{-1}$) for both the increment and minimum contour level; \citet{Williams94} also strongly recommend these values for the two parameters. We show an example of the clouds extracted by the algorithm in several subsequent velocity channels in Figure~\ref{cloud1}.

All clumps returned by the algorithm have integrated flux measurements greater than 3$\sigma$ (0.33 Jy beam$^{-1}$). Additionally, the velocity profile of each clump is visually inspected using the $\it{clplot}$ package within $\sc{CLUMPFIND}$, and any clumps with profiles which do not appear to be single entities (i.e., apparently blended profiles) are removed from the sample. 

The algorithm returns the FWHM extents of each clump in the x-, y-, and v-directions, calculates the total two-dimensional areas of each clump from the pixels assigned and the total detected flux within the three dimensional clump volume, and derives clump radii assuming the calculated areas are circular. We account for the beam size (the spatial resolution element) by subtracting in quadrature the beam size of a point source with the same peak temperature as the clump, measured out to the radius at which the increment equals the minimum contour level (where T=$\Delta$T), from the radii calculated using the area method (as described in Appendix A of \citealt{Williams94}). We account for the velocity resolution, or the (non-Gaussian) channel width, in our velocity dispersions by approximating the instrumental dispersion to be half of the channel width and subtracting this value in quadrature from the measured velocity dispersions. The equations are:

\begin{equation} 
R = \sqrt { R_{meas}^{2} -  \bigg( \frac{b}{2.355} \sqrt{2 ln \big( \frac{T_{peak}}{\Delta T} \big) } \bigg) ^{2} }
\end{equation}
and \\
\begin{equation} 
\sigma_{v} =  \sqrt { \sigma_{v, meas}^{2} - \big( \frac{\Delta v_{chan}}{2} \big) ^{2} }. 
\end{equation}

\citet{Rosolowsky06} define more complex treatments of these quantities which require the extrapolations of clouds out to 0 K. We find good agreement between the values for the velocity dispersion (to within 4\%) calculated using both methods, but the size measurements diverge for clouds smaller than our beam size. Since our measurements of size and velocity dispersion are measured out to a contour equal to twice the rms, not extrapolated to 0 K, we opt to use the Williams method to estimate the sizes and velocity dispersions of the clouds.

Following e.g., \citet{Wilson90}, we take the uncertainties in our measurements to be 25\% of the spatial beam size (17 pc) and half of our velocity channel width (1.3 \kms). These values are appropriate for the largest clouds in our sample, but the uncertainties in the clouds with sizes and velocity dispersions near our resolution limits may be larger. Running $\sc{CLUMPFIND}$ to a lower minimum contour level (1.5$\sigma$) creates more detections of faint clouds, but the resulting change in $X_{\text{CO}}$ within the individual GMCs presented in this paper is well within these quoted measurement errors.

\subsubsection{Derived properties}

We use the cloud measurements to calculate cloud virial masses in the following manner. Using the definition of potential energy from \citet{McKee92} and the kinetic energy in terms of the one-dimensional velocity dispersion, \citet{Williams94} express the virial mass as

\begin{equation}
M_{vir} = \frac{ 5 \Delta R \sigma_{v}^{2} } { \alpha_{vir} G }, 
\end{equation}

\noindent where $\Delta$R is the circular radius of a clump, $\sigma_{v}$ is the one-dimensional velocity dispersion, G is the gravitational constant, and $\alpha_{vir}$ is the ``virial parameter" which describes a non-uniform density profile. If we parameterize the density profile as 

\begin{equation} 
\rho(r) = \rho_{1} \bigg( \frac{R_1}{r} \bigg)^{\beta}, 
\end{equation}

\noindent we find that integrating over $\it{dM = \rho(r) y dy dx}$ yields 

\begin{equation} 
M_{vir} = \frac{ 4 \pi \rho_{1} R_{1}^{3} }{ 3 - \beta}.
\end{equation}

Substituting the expressions in Equations 5 and 6 into the gravitational energy for a spherical body (Equation 2.32, \citealt{BinneyTremaine08}) yields the relation 

\begin{equation}
\frac{3}{5} \alpha_{vir} = \frac{ 3 - \beta }{ 5 - 2\beta }. 
\end{equation}

It follows that for $\beta$=0, $\alpha_{vir}$=1; for $\beta$=1, $\alpha_{vir}$=10/9; and for $\beta$=2, $\alpha_{vir}$=5/3 (as also mentioned by \citealt{Williams94}). 

If we assume that $\beta$=1, as many authors do, (e.g., \citealt{Solomon87, Rosolowsky06, Bolatto08}) and G = 1/232 (using units of \kms, parsecs, and solar mass), substituting into Equation 4 yields

\begin{equation}
M_{vir} = 1040~R~\sigma_{v}^{2}, 
\end{equation}

\noindent as shown by \citet{Solomon87} and similarly by \citet{Wilson90}. We assume that the GMCs are virialized, an assumption consistent with previous studies, and derive virial masses using the measurements of R and $\sigma_{v}$ (corrected for their respective resolution elements) via Equation 8.

\citet{Solomon87} describe the virial mass as

\begin{equation}
M_{vir} = \frac{ 3 f_{p} S \sigma_{v}^{2} } { G }, 
\end{equation}

\noindent where f$_{p}$ is called a projection factor and the size of the cloud (S) is related to the effective radius (derived from the circular area of the cloud on the sky) such that R$_{eff}$ = 1.91 S. However, this projection factor, indicated to equal 2.9, is not explicitly derived and as a result does not appear in many recent papers on this topic. Using the formulae shown above, it is trivial to derive that f$_{p}$=2.9 when $\alpha_{vir}$=10/9. 

To calculate $X_{\text{CO}}$, in effect the mass-to-light ratio for molecular clouds, we compare the virial masses of each cloud to the luminosities derived from each cloud's integrated CO flux. We calculate cloud luminosities via

\begin{equation} 
L_{CO} = \frac{ 13.6 \lambda_{mm}^{2} F_{CO} } { \theta_a \theta_b }, 
\end{equation}

\noindent where $\lambda$ is the observed wavelength (in mm), F$_{CO}$ is the flux density in Jy beam$^{-1}$ \kms arcsec$^{2}$, and $\theta_{a}$ and $\theta_{b}$ are the beam axes (arcsec). Finally, with M$_{vir}$ in units of solar masses and L$_{CO}$ in units of K \kms pc$^{2}$, and including a factor of 1.36 to account for helium, we compute $X_{\text{CO}}$ (the CO-to-H$_{2}$ conversion factor) as follows:

\begin{equation} 
X_{\text{CO}} [\text{cm}^{-2}~\text{(K~km~s}^{-1})^{-1}] = 4.60 \times 10^{19}~\frac{ M_{vir} } { L_{CO} }.
\end{equation}



\section{Discussion}

\subsection{Structure and dynamics of CO}

From Figure~\ref{co}, it is clear that most of the bright CO emission which is included in the resolved GMCs is coincident with spiral arms, but significant emission is also found in the inter-arm regions. The GMCs identified by the $\sc{CLUMPFIND}$ algorithm tend to be the largest, brightest clouds: those in the spiral arms. The inter-arm emission appears to be more extended, which is consistent with it being less likely to display an apparently self-gravitating velocity profile. 


\subsection{Properties of individual GMCs}

We resolve a total of 134 clouds. After discarding potential blends, as described in Section 3.2, 120 clouds remain in our GMC sample. Using the equations in Section 3.2 to account for the instrumental resolution elements, 64 clouds have a real velocity dispersion. We require that clouds have a velocity dispersion of at least 2.54 \kms (twice our estimated instrumental dispersion) to be included in the sample, leaving 30 resolved clouds. 
The entire sample of 134 identified clouds is useful to examine the overall distribution of molecular gas which exists in GMCs, but the 30 fully resolved clouds will be the ones for which we derive virial masses, luminosities, and $X_{\text{CO}}$. These clouds are listed in Table~\ref{table:clouds}.

\begin{figure}
\plotone{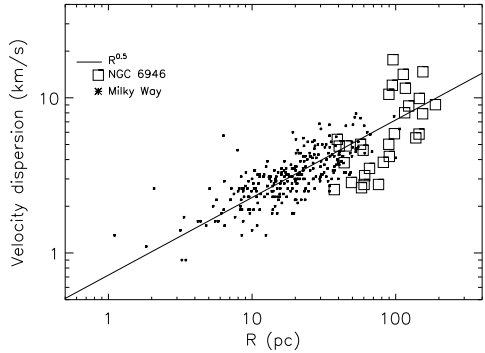}
\caption{The radii plotted against the velocity dispersions of the GMC sample detected in NGC 6946 are shown. The properties of the Milky Way GMC sample measured by Solomon et al. (1987) are plotted for comparison. 
The empirical relation of Solomon et al. of 0.72 $\times$ R$^{0.5}$ is also shown. 
\label{rsigma} }
\end{figure}

The radii and velocity dispersions of the 30 fully resolved GMCs are shown in Figure~\ref{rsigma}, as are the measurements of resolved Milky Way disk clouds detected by \citet{Solomon87} for comparison. \citet{Solomon87} performed the seminal study of GMCs in the Galactic disk and showed that the relationship between the sizes and velocity dispersions of GMCs are not related by an exponent of $\frac{1}{3}$, as predicted by Kolmogorov turbulence \citep{Larson81}, but instead by an index of 0.5, as is consistent with clouds in virial equilibrium. 

However, more recently, comparisons of GMCs at the Galactic center using various tracers have been made to those throughout the disk, indicating that departures from the traditional relation do occur for clouds at the Galactic center \citep{Miyazaki00, Oka01}.
The observed tendency is for clouds to exhibit higher velocity dispersions for a given size. Although our map covers the central region of NGC 6946, the measurements of its GMCs are largely consistent with the Galactic disk relation with some scatter, particularly for the largest clouds. The measurements of radius and velocity dispersion may even be consistent with a slope larger than 0.5, the value measured by several authors for Milky Way clouds within the disk and at the Galactic center \citep{Solomon87, Miyazaki00, Oka01}. 

For the largest clouds, the observed scatter goes to higher $\sigma_{v}$ for a given size, and in fact, the six clouds which fall well above the Solomon relation are all located within $\sim$400~pc of the center of NGC 6946, consistent with the trend observed to be true for the center of our Galaxy. The smallest clouds which deviate from the Solomon relation are very close to our velocity resolution limit. While we are not sensitive to the full range of GMCs detected in the Galaxy by \citet{Solomon87}, our detected GMCs are largely consistent with the biggest clouds in the Galactic sample. While the Milky Way clouds included in the \citet{Solomon87} study were measured using a contouring method which is physically distinct from the decomposition algorithm employed in this paper, the consistent trends present in both samples indicate similar underlying physics at work in both the center of NGC 6946 and the disk of our Galaxy.

\begin{figure}
\plotone{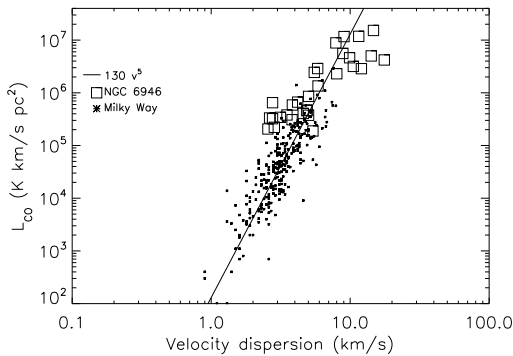}
\caption{Luminosities and velocity dispersions of the NGC 6946 and Galactic \citet{Solomon87} samples are shown. The fit to the Galactic points, L$_{CO}$ $\approx$ 130 $\sigma_{v}^{5}$ K \kms pc$^{2}$, is also shown.  
\label{sigma_lco} }
\end{figure}

The GMC luminosities are shown in Figure~\ref{sigma_lco}. The Galactic points and their linear fit \citep{Solomon87} are also shown for comparison. Again, our GMC sample is largely consistent with the measurements of the brightest Galactic clouds in the Solomon sample, though a fit to the NGC 6946 clouds would be shallower than the relation measured by \citet{Solomon87}. The brightest GMCs -- the same clouds which fall above the Solomon radius-$\sigma_{v}$ relation -- appear to be underluminous for their velocity dispersion compared to the Galactic relation. Though blended clouds could artificially produce this relationship, since the total luminosity of multiple optically thick $^{12}$CO clouds along the same line of sight may not increase linearly with the amount of CO-emitting gas (depending on the geometry), the velocity dispersion would betray the presence of more than one cloud via multiple peaks. We individually inspect each cloud to select them on the basis of their apparently unblended velocity profiles, so blended clouds (at least at our velocity resolution limit) can not explain this finding. 


\subsection{CO-to-H$_{2}$ conversion factor}

\begin{figure}
\plotone{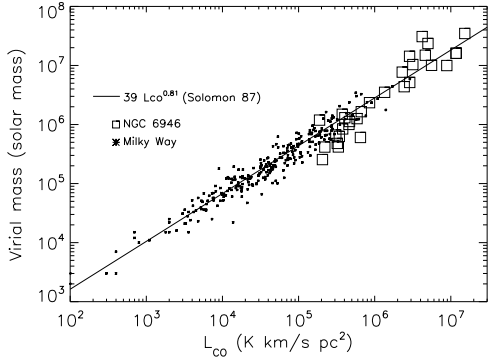}
\caption{Luminosities and virial masses of the NGC 6946 GMC sample and the Galactic \citet{Solomon87} sample are shown. The overplotted fit to the Galactic points, M$_{vir}$ = 39 $\times$ (L$_{CO}$)$^{0.81}$, is consistent with both sets of GMCs. \label{lco_mvir} }
\end{figure}

\begin{figure}
\plotone{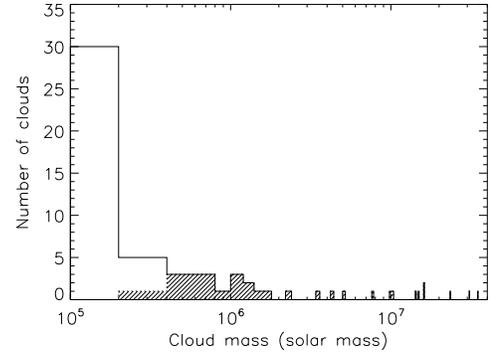}
\caption{Histogram of the GMC masses in the NGC 6946 sample. The solid line indicates the sample of 120 resolved clouds with apparently self-gravitating profiles, and the shaded area indicates the sample of 30 clouds on which we base our analysis. \label{hists} }
\end{figure}

The virial masses and CO luminosities of our NGC 6946 GMC sample are shown in Figure~\ref{lco_mvir}. Once again, we compare our measurements to the Galactic sample; the overlaid linear fit is derived in \citet{Solomon87}. The median mass of our GMC sample is comparable to the most massive Galactic disk GMCs, but we find excellent consistency between the trends. The histogram of our GMC sample is shown in Figure~\ref{hists}. 


  
Including a factor of 1.36 to account for heavier elements, the average value of $X_{\text{CO}}$ for the GMC sample is 1.2 $\times$ 10$^{20}$ cm$^{-2}$ (K \kms)$^{-1}$; this value can vary by as much as a factor of 2 with our assumed uncertainties. This value is roughly a factor of two below the average value of the conversion factor which we calculate from the \citet{Solomon87} Galactic disk clouds of 2.6 $\times$ 10$^{20}$ cm$^{-2}$ (K \kms)$^{-1}$ (derived using the same factor for heavier elements). If we calculate $X_{\text{CO}}$ from the \citet{Solomon87} clouds using only GMCs more massive than 5 $\times$ 10$^{5}$ \Msun -- approximately our completeness limit -- we find a value of 1.9 $\times$ 10$^{20}$ cm$^{-2}$ (K \kms)$^{-1}$, which is in agreement with the value we derive for NGC 6946 within our uncertainty. It is also in agreement with the value derived by \citet{Dame01} of 1.8 ($\pm$0.3) $\times$ 10$^{20}$ cm$^{-2}$ (K \kms)$^{-1}$ for CO emission in the solar neighborhood.

In addition to CO measurements, the conversion factor in the Galaxy has also been derived using $\gamma$-ray measurements \citep{Strong96}, which yield a conversion factor of 1.9 $\times$ 10$^{20}$ cm$^{-2}$ (K \kms)$^{-1}$, in good agreement with the values listed above. This agreement between values of $X_{\text{CO}}$ determined using both dynamical and non-dynamical methods has been claimed as evidence that Galactic GMCs are gravitationally bound and in equilibrium. \citet{Heyer09} correct the \citet{Solomon87} measurements for a more recently derived Galactic center distance and a factor of 1.36 to account for heavier elements and derive a mean surface density of 206 M$_\sun$ pc$^{-2}$ for the entire Galactic disk GMC sample; the mean surface density that we calculate within our clouds is 204 \Msun pc$^{-2}$.

\begin{figure}
\plotone{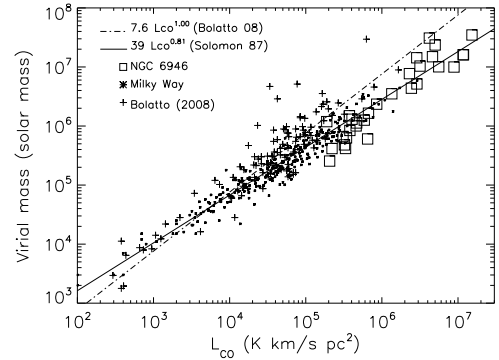}
\caption{The same as Figure~\ref{lco_mvir}. Extragalactic GMC properties from the study by \citet{Bolatto08}, and their corresponding fit, are shown as well as the \citet{Solomon87} fit to the Milky Way sample. The clouds measured in NGC 6946 are more consistent with the fit to the Galactic GMCs. \label{lco_mvir_withbol} }
\end{figure}

The scatter in measurements of the conversion factor among studies of nearby galaxies has tended historically to be large, but more recently, studies of $X_{\text{CO}}$ have exhibited relative agreement with the Galactic value. In \citet{Bolatto08}, $^{12}$CO (J=1-0) and (J=2-1) measurements of GMCs within Local Group dwarf galaxies and spiral galaxies are compiled and compared to the molecular clouds detected in the Milky Way. The sample includes dwarf galaxies as well as the Large and Small Magellanic Clouds (LMC and SMC) to investigate the effect played by the environment on determinations of $X_{\text{CO}}$. 
These authors find that GMCs in Local Group dwarfs obey similar relationships to Milky Way GMCs and are consistent, to within a factor of two, with an $X_{\text{CO}}$ of 2 $\times$ 10$^{20}$ cm$^{-2}$ (K \kms)$^{-1}$ with no detectable dependence on metallicity. A comparison of the GMC masses and luminosities of the NGC 6946 sample and the extragalactic GMC sample from \citet{Bolatto08} is presented in Figure~\ref{lco_mvir_withbol}. \citet{Blitz07} also study GMCs within nearby galaxies and find that $X_{\text{CO}}$ is within a factor of two of 4 $\times$ 10$^{20}$ cm$^{-2}$ (K \kms)$^{-1}$. Even GMCs in the outer disk of M33 exhibit characteristics similar to those of Galactic GMCs \citep{Bigiel2010}. In both the studies by \citet{Bolatto08} and by \citet{Blitz07}, however, the low-metallicity SMC is an outlier; in \citet{Blitz07}, it requires a conversion factor three times higher than the rest of the GMCs in the sample.

As it is common for disk galaxies to exhibit a metallicity gradient, we investigate whether a radial dependence of the distributions of sizes, masses, and $X_{\text{CO}}$ values of individual GMCs exists. Various observational studies which correlate the conversion factor with metallicity do so by studying them as a function of radius (or even averaged over entire disks) to investigate how $X_{\text{CO}}$ may vary with metallicity in the presence of a metallicity gradient, especially in diffuse and low-metallicity gas \citep{Arimoto96, Leroy11, Genzel2011}. For instance, \citet{Nakai95} show that M51 exhibits a radial gradient in the conversion factor in M51, with $X_{\text{CO}}$ increasing with radius as the metallicity decreases. The same has also been shown to be true in the Galaxy \citep{Sodroski91}. 

\begin{figure*}
\plottwo{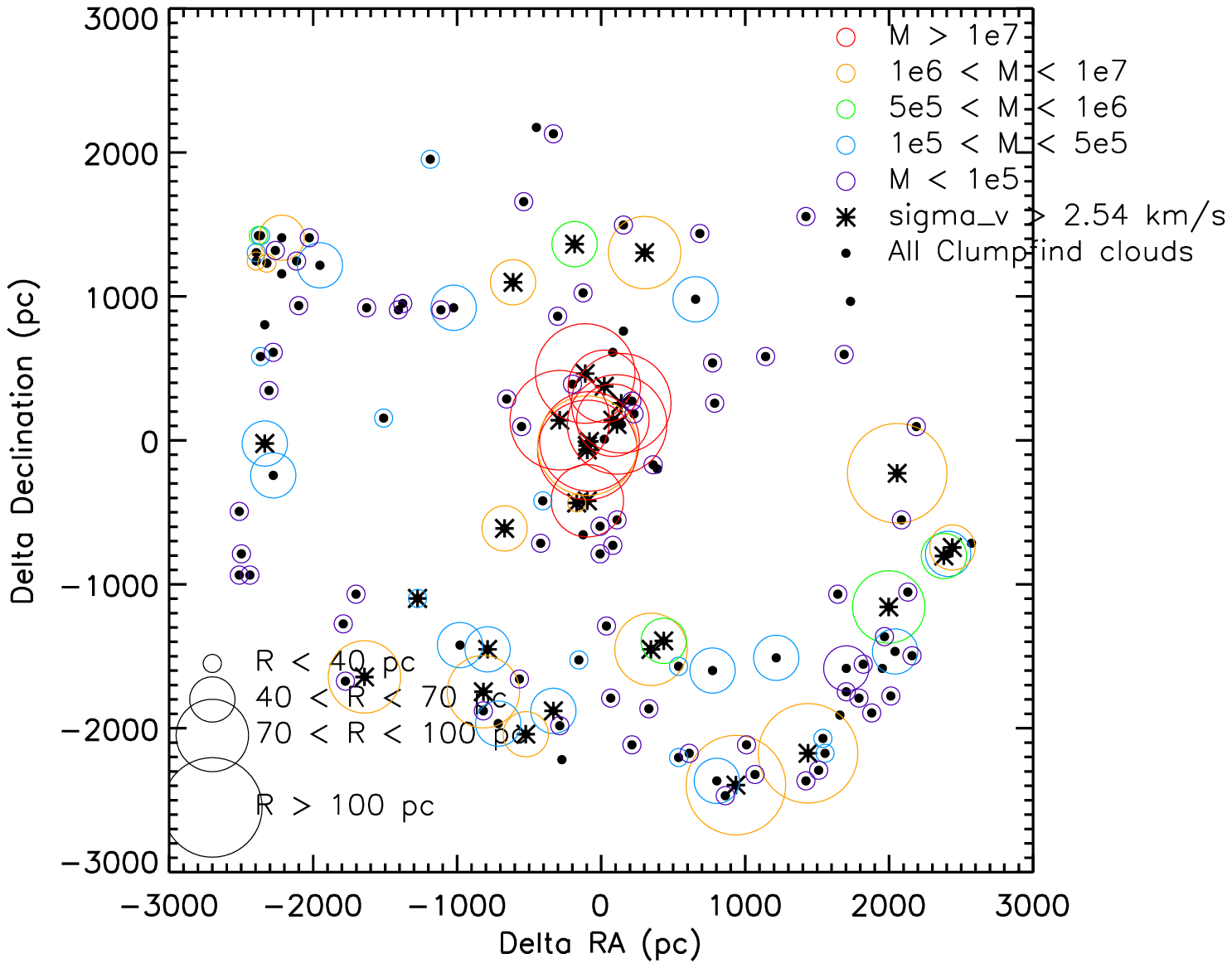}{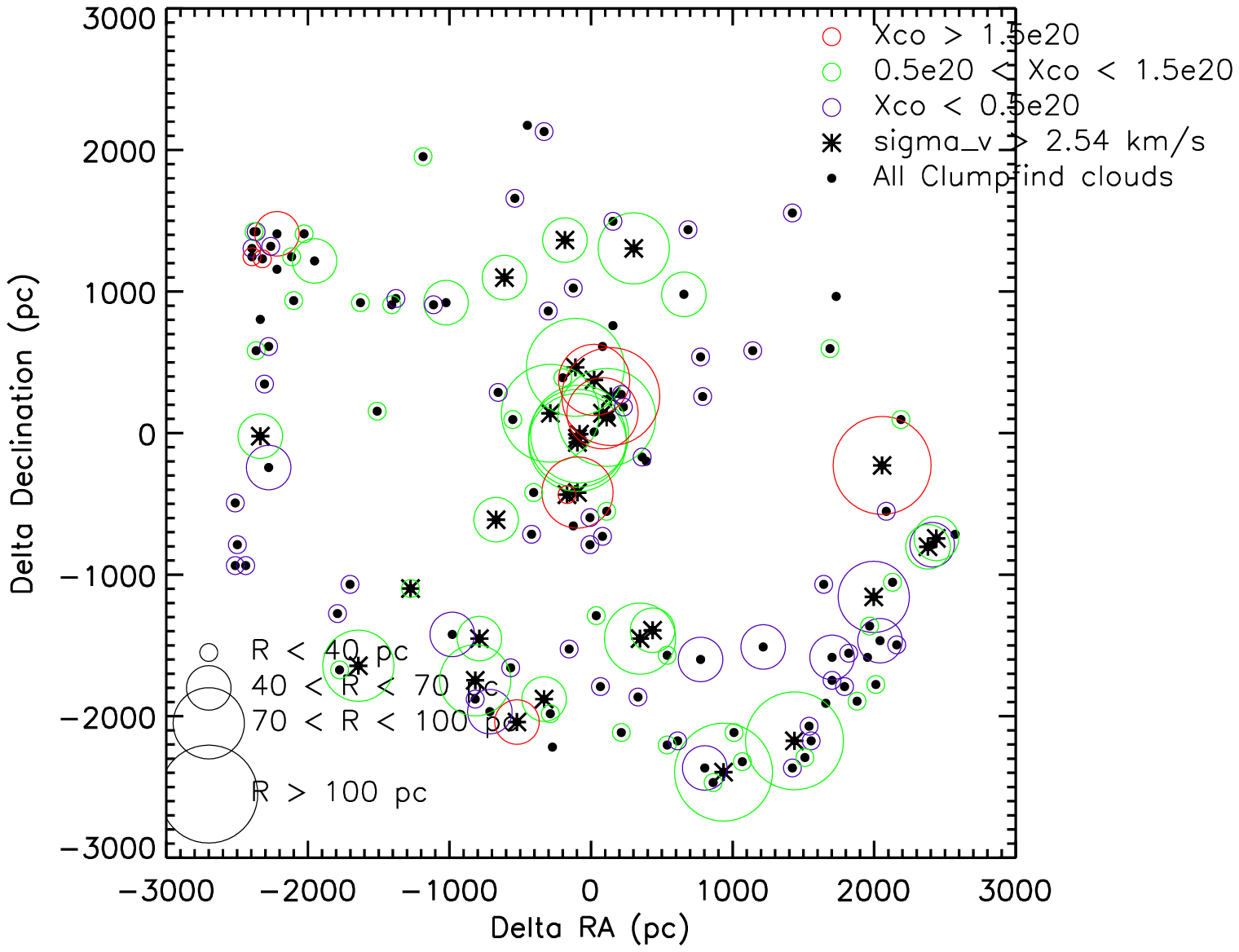}
\caption{Spatial distribution of radii, masses, and $X_{\text{CO}}$ values. In these plots, all 134 resolved clouds are shown as dots. The 120 clouds whose velocity profiles indicate that the clouds are apparently self-gravitating entities are identified by a colored circle as well. The 30 clouds which have real velocity dispersions after subtracting the velocity resolution element, which used in our analysis, are indicated by asterisks. The radii, mass, and $X_{\text{CO}}$ of each cloud with such a velocity profile is coded by color and size to indicate the values of these properties. \label{maps} }
\end{figure*}

These properties are displayed in Figure~\ref{maps}. In this figure, the full detected cloud sample of 134 clouds is shown as dots. The 120 clouds with velocity profiles which point unambiguously to self-gravitation are circled in a color which indicates the mass (left) or value of $X_{\text{CO}}$ (right) while the size of the circles indicates the radius of each cloud. The 30 clouds used in our analysis are indicated by asterisks. The most massive clouds ($>$ 10$^{7}$ M$_{\odot}$) are found at the very center of NGC 6946, as described in Section 4.2, and clouds with masses $>$ 5 $\times$ 10$^{5}$ M$_{\odot}$ are found where the density of clouds is the highest (i.e., on the spiral arms). No radial trend is apparent in $X_{\text{CO}}$, though the most central clouds tend to have values higher than 0.5 $\times$ 10$^{20}$ cm$^{-2}$ (K \kms)$^{-1}$, and all but one of the clouds with $X_{\text{CO}}$ below this value are smaller than 70~pc. 

\subsubsection{Metallicity}

At radii less than 2.5~kpc from the center of NGC 6946, the mean O/H abundance level is roughly constant \citep{Belley92}; this is the region over which we present resolved measurements of GMCs, so we do not expect systematic or radial changes in metallicity to affect our measured values of $X_{\text{CO}}$. However, it is still worthwhile to compare our derived value of the conversion factor to the metallicity at the center of NGC 6946 in the context of previously derived trends relating the two. Typically, the conversion factor is found to increase as the host galaxy metallicity decreases. The metallicity measured at the center of NGC 6946 quoted by \citet{Moustakas2010} is [12 + log (O/H)] = 8.47$\pm$0.09 or [12 + log (O/H)] = 9.16$\pm$0.06, depending on the empirical or theoretical abundance calibration applied. The two values are intended to indicate the range of metallicities which would be derived using current strong-line calibrations \citep{Moustakas2010}.

Our derived value of $X_{\text{CO}}$ and the lower Moustakas value of the metallicity are consistent with the relation $\alpha$/$\alpha_{Gal}$ = 6.0 -- 0.67[12 + log (O/H)] found by \citet{Wilson95}, where the correlation between $X_{\text{CO}}$ and metallicity is derived using virial GMC measurements. In addition to using virial measurements, the metallicity dependence of $X_{\text{CO}}$ has also been tested by assuming that dust traces gas (i.e., \citealt{Magrini2011}). Our derived value of $X_{\text{CO}}$ and the higher Moustakas value of the metallicity are consistent with the trends found using this method in \citet{Israel00} and \citet{Boselli02}, which are shown over 2-3 orders of magnitude in $X_{\text{CO}}$. On the other hand, \citet{Leroy11} find that metallicity is correlated with the conversion factor at metallicities below [12 + log (O/H)] = 8.2; the entire Moustakas range of values for the metallicity at the center of NGC 6946 is higher than this value.  

In the Milky Way, \citet{Balser2011} measure a radial metallicity gradient in a sample of Galactic disk HII regions to be [12 + log (O/H)] = (8.962 $\pm$ 0.045) -- (0.0446 $\pm$ 0.0046) $\it{R_{gal}}$, which yields Galactic disk metallicities in the range of 8.5 to 8.7 at radii of 5-10~kpc. These measurements of the Galactic disk metallicity are consistent with the Moustakas range of values of the metallicity at the center of NGC 6946; thus, it is not unreasonable to expect that the GMCs in the two regions are physically similar, especially within our quoted uncertainty for $X_{\text{CO}}$.

\subsection{``Diffuse" fraction of CO}

The fraction of emission in our CO cube which is identified as emission from the 30 GMCs with approximately Gaussian velocity profiles (suggesting self-gravitation) greater than our velocity resolution element is 19\%; the total mass in this component is 2.2$\times$10$^{8}$ \Msun. If we include all of the GMCs identified by $\sc{CLUMPFIND}$, which includes clouds which are spatially resolved but exhibit blended velocity profiles, the fraction increases to 31\%. However, since the algorithm discards emission contours below 2$\sigma$, fainter emission around GMCs is not included. While this does not affect $X_{\text{CO}}$, as described in Section 3.2, it does artificially elevate the ``diffuse" fraction that we find. For instance, the fraction of emission included in all GMCs in the 1.5$\sigma$ run is 47\% (compared to 31\%). 

The critical density for collisional excitation of CO (J=1-0) emission is 300 cm$^{-3}$ \citep{ScovilleSanders87}, which is roughly the average gas density within a GMC \citep{Solomon87}. The CO emission that is not identified as belonging to GMCs in our analysis is unlikely to be entirely diffuse (i.e., below the critical density). In addition, our velocity resolution of 2.54 \kms -- optimal for resolving more massive GMCs -- is not optimal for the detection of less massive GMCs.

We expect to be sensitive to clouds down to a limit of $\sim$5 $\times$ 10$^{5}$ M$_{\odot}$, but certainly clouds less massive than this limit will contribute as well. In their Galactic GMC sample, \citet{ScovilleSanders87} find that $\sim$50\% of the total H$_{2}$ mass resides in clouds less massive than 4 $\times$ 10$^{5}$ M$_{\odot}$, and the other half is found in clouds more massive than this value. Taking our estimate -- that roughly half of the CO emission that we detect (in our 1.5$\sigma$ run of $\sc{CLUMPFIND}$) is assigned to clouds more massive than $\sim$5 $\times$ 10$^{5}$ M$_{\odot}$ -- in conjunction with the \citet{ScovilleSanders87} result for the Galaxy indicates that the fraction of truly diffuse CO that we detect is likely small. 

\section{Conclusions}

In this paper, we present the first galaxy from our CO Survey of Nearby Galaxies taken with CARMA and NRO45. The high image fidelity afforded by our combined observations allows us to study extragalactic molecular clouds with unprecedented resolution. We utilize the clump-finding algorithm $\sc{CLUMPFIND}$ \citep{Williams94} with the recommended rms noise-dependent inputs and find that the general trends recovered by the clouds resemble those of Galactic clouds \citep{Solomon87}, but the clouds within $\sim$400~pc of the center of NGC 6946 exhibit higher velocity dispersion for a given size, similar to the behavior observed by \citet{Oka01} for clouds at the Galactic center. These clouds also exhibit higher velocity dispersions for a given luminosity, which is not likely to be caused by blends of clouds (at least at our velocity resolution limit). 

We derive an average value of $X_{\text{CO}}$ of 1.2 $\times$ 10$^{20}$ cm$^{-2}$ (K \kms)$^{-1}$, which can vary by as much as a factor of 2 with our assumed measurement uncertainties, and is consistent within our errors with accepted Galactic values of $X_{\text{CO}}$ \citep{Dame01}. The trends which we observe among the GMCs at the center of NGC 6946 are broadly consistent with Galactic trends seen by \citet{Solomon87}, even though we use a phenomenologically different algorithm to define the clouds, and we detect clouds comparable to the brightest and most massive clouds in the Galactic sample. The most massive clouds ($>$ 10$^{7}$ M$_{\odot}$) are found at the very center of NGC 6946, and clouds with masses $>$ 5 $\times$ 10$^{5}$ M$_{\odot}$ are found where the density of clouds is the highest (i.e., on the spiral arms). No radial trend is apparent in $X_{\text{CO}}$, though the most central clouds tend to have values higher than 0.5 $\times$ 10$^{20}$ cm$^{-2}$ (K \kms)$^{-1}$, and all but one of the clouds with $X_{\text{CO}}$ below this value are smaller than 70~pc.

When we run the clump decomposition algorithm down to 1.5$\sigma$, which does not change our measured value of $X_{\text{CO}}$ but does incorporate more extended emission around our detected GMCs, we find that 47\% of our detected CO emission is identified as belonging to a GMC. Our study is optimized to resolve larger and more massive GMCs as opposed to detecting less massive GMCs; we are most sensitive to clouds above $\sim$5 $\times$ 10$^{5}$ M$_{\odot}$. If we adopt the \citet{ScovilleSanders87} finding in the Galaxy that 50\% of the H$_{2}$ mass is in clouds more massive than 4 $\times$ 10$^{5}$ M$_{\odot}$ and 50\% is in clouds less massive than this value, then the fraction of truly diffuse CO emission that we detect in NGC 6946 must be small. 

Studies of this nature will soon be expanded when ALMA comes online, as observations will be feasible with better sensitivity and resolution in a fraction of the observing time.

\acknowledgments
The authors would like to thank the anonymous referee for the suggestions which improved the discussion of this paper. This research has made use of the NASA/IPAC Extragalactic Database (NED) which is operated by the Jet Propulsion Laboratory, California Institute of Technology, under contract with the National Aeronautics and Space Administration.

\begin{center}
\begin{table*}
\caption{Properties of GMCs in NGC 6946  \label{table:clouds}}
\medskip
\begin{tabular}{lccccccc}
\hline
\hline
Number & RA$_{J2000}$ & Dec$_{J2000}$ &   R (pc) & $\sigma_{v}$ (\kms) & L$_{CO}$ (10$^{5}$ K \kms pc$^{2}$) & M$_{vir}$ (10$^{5}$ M$_{\odot}$) & $X_{\text{CO}}$ (10$^{20}$ [cm$^{-2}$ (K \kms)$^{-1}$]) \\
\hline
       1 &         20$\h$     34$\m$      52.73$\s$&       60$\dg$     9$\am$     12$\as$ &        117 &        11.5 &        118 &        161 &       0.630 \\
       2 &         20$\h$     34$\m$      52.66$\s$&       60$\dg$     9$\am$     14$\as$ &        155 &        14.7 &        152 &        348 &        1.05 \\
       3 &         20$\h$     34$\m$      52.73$\s$&       60$\dg$     9$\am$     13$\as$ &        154 &        7.89 &        88.5 &        99.7 &       0.518 \\
       4 &         20$\h$     34$\m$      53.60$\s$&       60$\dg$     9$\am$     19$\as$ &        189 &        9.04 &        117 &        161 &       0.633 \\
       5 &         20$\h$     34$\m$      51.79$\s$&       60$\dg$     9$\am$     18$\as$ &        123 &        8.86 &        55.8 &        101 &       0.829 \\
       6 &         20$\h$     34$\m$      52.73$\s$&       60$\dg$     9$\am$    &        89.7 &        10.5 &        31.7 &        104 &        1.50 \\
       7 &         20$\h$     34$\m$      52.80$\s$&       60$\dg$     9$\am$     30$\as$ &        146 &        9.91 &        46.2 &        149 &        1.49 \\
       8 &         20$\h$     34$\m$      48.05$\s$&       60$\dg$     7$\am$     53$\as$ &        145 &        5.84 &        28.9 &        51.6 &       0.822 \\
       9 &         20$\h$     34$\m$      51.66$\s$&       60$\dg$     9$\am$     23$\as$ &        113 &        14.2 &        50.1 &        236 &        2.17 \\
      10 &        20$\h$     34$\m$      41.48$\s$&       60$\dg$     8$\am$     47$\as$ &        43.8 &        3.82 &        3.09 &        6.64 &       0.989 \\
      11 &        20$\h$     34$\m$      52.20$\s$&       60$\dg$     9$\am$     27$\as$ &        94.8 &        12.0 &        28.7 &        143 &        2.29 \\
      12 &        20$\h$     34$\m$      51.93$\s$&       60$\dg$     9$\am$     19$\as$ &        95.8 &        17.6 &        42.0 &        309 &        3.38 \\
      13 &        20$\h$     34$\m$      45.77$\s$&       60$\dg$     8$\am$     &        138 &        5.52 &        24.4 &        43.7 &       0.825 \\
      14 &        20$\h$     34$\m$      41.21$\s$&       60$\dg$     8$\am$     49$\as$ &        40.4 &        4.89 &        4.59 &        10.0 &        1.01 \\
      15 &        20$\h$     34$\m$      42.95$\s$&       60$\dg$     9$\am$     6.5$\as$ &        116 &        8.01 &        23.0 &        77.1 &        1.54 \\
      16 &        20$\h$     34$\m$      50.72$\s$&       60$\dg$     8$\am$     25$\as$ &        98.0 &        5.87 &        13.5 &        35.1 &        1.20 \\
      17 &        20$\h$     34$\m$      55.08$\s$&       60$\dg$     9$\am$     51$\as$ &        45.2 &        4.90 &        4.66 &        11.3 &        1.11 \\
      18 &        20$\h$     34$\m$      43.22$\s$&       60$\dg$     8$\am$     35$\as$ &        76.0 &        2.76 &        6.49 &        6.04 &       0.428 \\
      19 &        20$\h$     34$\m$      59.76$\s$&       60$\dg$     8$\am$     18$\as$ &        89.5 &        5.02 &        8.55 &        23.4 &        1.26 \\
      20 &        20$\h$     35$\m$      2.913$\s$&       60$\dg$     9$\am$     13$\as$ &        57.8 &        2.63 &        3.33 &        4.17 &       0.576 \\
      21 &        20$\h$     34$\m$      55.88$\s$&       60$\dg$     8$\am$     25$\as$ &        60.3 &        2.76 &        3.22 &        4.79 &       0.683 \\
      22 &        20$\h$     34$\m$      56.01$\s$&       60$\dg$     8$\am$     15$\as$ &        82.0 &        3.84 &        5.87 &        12.6 &       0.986 \\
      23 &        20$\h$     34$\m$      58.09$\s$&       60$\dg$     8$\am$     37$\as$ &        37.4 &        2.56 &        2.06 &        2.55 &       0.569 \\
      24 &        20$\h$     34$\m$      53.80$\s$&       60$\dg$     8$\am$     10$\as$ &        49.3 &        2.85 &        2.21 &        4.15 &       0.863 \\
      25 &        20$\h$     34$\m$      55.34$\s$&       60$\dg$     8$\am$     53$\as$ &        59.3 &        4.59 &        4.06 &        13.0 &        1.48 \\
      26 &        20$\h$     34$\m$      54.67$\s$&       60$\dg$     8$\am$     5.0$\as$ &        56.9 &        5.02 &        3.75 &        14.9 &        1.83 \\
      27 &        20$\h$     34$\m$      50.92$\s$&       60$\dg$     9$\am$     58$\as$ &        90.3 &        4.19 &        6.66 &        16.5 &        1.14 \\
      28 &        20$\h$     34$\m$      53.13$\s$&       60$\dg$    10$\am$    &        65.7 &        3.51 &        3.80 &        8.42 &        1.02 \\
      29 &        20$\h$     34$\m$      50.32$\s$&       60$\dg$     8$\am$     27$\as$ &        60.9 &        3.15 &        3.43 &        6.29 &       0.844 \\
      30 &        20$\h$     34$\m$      53.07$\s$&       60$\dg$     8$\am$     59$\as$  &        39.2 &        5.40 &        1.89 &        11.9 &        2.90 \\
      

\hline
\end{tabular}
\end{table*}
\end{center}

\bibliography{jen_refs}

\end{document}